\documentclass[
amsmath,
amssymb,
amsfonts,
aps,
nofootinbib,
preprintnumbers,
prl,
reprint,
superscriptaddress
]{revtex4-2}

\usepackage{xcolor}
\definecolor{MyDarkBlue}{rgb}{0.15,0.25,0.45} 
\usepackage[
linktocpage=true,
hypertexnames=false,
colorlinks=true,
citecolor=MyDarkBlue,
linkcolor=MyDarkBlue,
urlcolor=MyDarkBlue,
pdfauthor={
    Leron Borsten,
    Branislav Jurco,
    Hyungrok Kim,
    Tommaso Macrelli,
    Christian Saemann,
    Martin Wolf},
pdftitle={Kinematic algebras from twistor space},
pdfsubject={hep-th},
breaklinks=true
]{hyperref}

\usepackage{amsthm}
\usepackage{mathtools}
\usepackage{bbm}
\usepackage{bm}
\usepackage{mathrsfs}
\usepackage{tikz}
\usetikzlibrary{matrix,cd,arrows}
\usepackage{stmaryrd}
\usepackage{cleveref}
\usepackage{ifthen}
\newcommand{\makecommand}[3]{%
    \foreach \i in #3 {%
        \expandafter\xdef\csname #1\i\endcsname{\noexpand#2{\unexpanded\expandafter{\i}}}%
    }%
}
\newcommand{\latinalphabet}{A,a,B,b,C,c,d,D,E,e,F,f,G,g,H,h,I,i,J,j,K,k,L,l,M,m,N,n,O,o,P,p,Q,q,R,r,S,s,T,t,U,u,V,v,W,w,X,x,Y,y,Z,z}
\makecommand{I}{\mathbbm}{\latinalphabet}
\makecommand{I}{\mathbbm}{{CP}}
\makecommand{bf}{\mathbf}{\latinalphabet}
\makecommand{bm}{\bm}{\latinalphabet}
\makecommand{ca}{\mathcal}{\latinalphabet}
\makecommand{fr}{\mathfrak}{\latinalphabet}
\makecommand{rm}{\mathrm}{\latinalphabet}
\makecommand{sf}{\mathsf}{\latinalphabet}
\makecommand{sf}{\mathsf}{{id,SU,Spin}}
\makecommand{sc}{\mathscr}{\latinalphabet}

\newcommand{\dpar}{\partial}
\newcommand{\wave}{\mathop\square\nolimits}

\newcommand{\tr}{\mathrm{tr}} 

\newcommand{\parder}[2][]{%
    \ifthenelse{\equal{#1}{}}{%
        \frac{\partial}{\partial #2}%
    }{%
        \frac{\partial #1}{\partial #2}%
    }%
}
\newcommand{\delder}[2][]{%
    \ifthenelse{\equal{#1}{}}{%
        \frac{\delta}{\delta #2}%
    }{%
        \frac{\delta #1}{\delta #2}%
    }%
}

\newtheoremstyle{breaknodot}
{\topsep}{\topsep}%
{\itshape}{}%
{\bfseries}{}%
{0pt}{\thmname{#1}\thmnumber{ #2.}~\thmnote{ \normalfont(#3)}}%
\theoremstyle{breaknodot}

\newcommand{\BVbox}{\text{BV}^\BBox}

\let\oldblacksquare\blacksquare
\newcommand{\BBox}{{\textcolor{gray}\oldblacksquare}} 

\newcommand\cn\caN 

\begin{document}
    
    \preprint{DMUS--MP--22/23}
    \preprint{EMPG--22--22}
    \title{Kinematic Lie Algebras from Twistor Spaces}
    \author{Leron Borsten}
    \email[]{l.borsten@herts.ac.uk}
    \affiliation{Department of Physics, Astronomy, and Mathematics, University of Hertfordshire,\\Hatfield AL10 9AB, United Kingdom}
    \author{Branislav Jur{\v c}o}
    \email[]{branislav.jurco@gmail.com}
    \affiliation{Mathematical Institute, Faculty of Mathematics and Physics,  Charles University Prague,\\ Prague 186 75, Czech Republic}
    \author{Hyungrok Kim}
    \email[]{hk55@hw.ac.uk}
    \affiliation{Maxwell Institute for Mathematical Sciences, Department of Mathematics, Heriot-Watt University,\\ Edinburgh EH14 4AS, United Kingdom}
    \author{Tommaso Macrelli}
    \email[]{tmacrelli@phys.ethz.ch}
    \affiliation{Department of Physics, ETH Zurich,\\ 8093 Zurich, Switzerland}
    \author{Christian Saemann}
    \email[]{c.saemann@hw.ac.uk}
    \affiliation{Maxwell Institute for Mathematical Sciences, Department of Mathematics, Heriot-Watt University,\\ Edinburgh EH14 4AS, United Kingdom}
    \author{Martin Wolf}
    \email[]{m.wolf@surrey.ac.uk}
    \affiliation{School of Mathematics and Physics, University of Surrey,\\ Guildford GU2 7XH, United Kingdom}
    
    \date{\today}
    
    \hyphenation{ms-sd-YM}
    
    \begin{abstract}
        We analyze theories with color--kinematics duality from an algebraic perspective and find that any such theory has an underlying $\BVbox$-algebra, extending the ideas of Reiterer~\cite{Reiterer:2019dys}. Conversely, we show that any theory with a $\BVbox$-algebra  features a kinematic Lie algebra that controls interaction vertices, both on- and off-shell. We explain that the archetypal example of a theory with a $\BVbox$-algebra is Chern--Simons theory, for which the resulting kinematic Lie algebra is isomorphic to the Schouten--Nijenhuis algebra on multivector fields. The $\BVbox$-algebra implies the known color--kinematics duality of Chern--Simons theory. Similarly, we show that holomorphic and Cauchy--Riemann Chern--Simons theories come with $\BVbox$-algebras and that, on the appropriate twistor spaces, these theories organize and identify kinematic Lie algebras for self-dual and full Yang--Mills theories, as well as the currents of any field theory with a twistorial description. We show that this result extends to the loop level under certain assumptions.
    \end{abstract}
    
    \maketitle
    
    \section{Introduction and summary}
    
    Color--kinematics (CK) duality~\cite{Bern:2008qj,Bern:2010ue} (see reviews~\cite{Carrasco:2015iwa,Borsten:2020bgv,Bern:2019prr,Adamo:2022dcm, Bern:2022wqg}) is a surprising property of certain field theories, that allows for their scattering amplitudes to be split into a kinematic component or numerators and gauge Lie algebra or color numerators, such that the kinematic numerators mirror the algebraic properties of the color numerators. This was first observed for the tree-level amplitudes of Yang--Mills theory, but many other theories also exhibit CK-duality. CK-duality is the cornerstone of the double copy prescription~\cite{Bern:2008qj,Bern:2010ue}, which constructs gravity scattering amplitudes from a simple combination of pairs of corresponding kinematical numerators for Yang--Mills theory, suggesting deep, illuminating connections between the known fundamental theories of nature and providing cutting-edge predictions in gravitational-wave astronomy, see e.g.~\cite{Bern:2021dqo,Bern:2019crd,Bern:2019nnu}. 
    
    Concretely, a theory is CK-dual if its $n$-point amplitudes $\caA_n$ can be written as 
    \begin{equation}\label{eq:YM_amplitudes_parameterization}
        \caA_n\sim\sum_{\gamma\in \Gamma_{n}}\frac{\sfc_\gamma\sfn_\gamma}{d_\gamma},
    \end{equation}
    where $\Gamma_{n}$ denotes the set of cubic Feynman graphs with $n$ external lines, $d_\gamma$ are the product of $\frac{1}{p^2_\ell}$ over all propagator lines $\ell$, with $p_\ell$ their momenta, $\sfc_\gamma$ are the color numerators, i.e.~products of gauge Lie algebra structure constants, as prescribed by the diagram $\gamma$, and $\sfn_\gamma$ are the kinematic numerators, built from momenta and polarization tensors. Furthermore, $\sfc_\gamma$ and $\sfn_\gamma$ obey the same antisymmetry under the interchange of edges in $\gamma$, and $\sfc_{\gamma_1}+\sfc_{\gamma_2}+\sfc_{\gamma_3}=0$ implies $\sfn_{\gamma_1}+\sfn_{\gamma_2}+\sfn_{\gamma_3}=0$. CK-duality thus suggests the existence of a \emph{kinematic Lie algebra} (KLA) from which the $\sfn_\gamma$ are constructed. 
    
    The geometric and algebraic underpinnings of this KLA remain a central question~\cite{BjerrumBohr:2009rd,Stieberger:2009hq,Feng:2010my,BjerrumBohr:2010hn,Broedel:2011pd,Mafra:2011kj,Mafra:2012kh,Fu:2012uy, Broedel:2012rc, Broedel:2013tta,He:2015wgf, Bjerrum-Bohr:2016axv, Mizera:2019blq, Reiterer:2019dys,Chi:2021mio,Borsten:2023ned}. In the case of self-dual Yang--Mills (SDYM) theory, the KLA consists of area-preserving diffeomorphisms on $\IC^2$~\cite{Monteiro:2011pc}, see~\cite{Bjerrum-Bohr:2012kaa,Boels:2013bi,Monteiro:2013rya,Fu:2016plh,Monteiro:2022lwm}. A cubic Lagrangian for the nonlinear sigma model with Feynman rules obeying $(\le1)$-loop CK-duality~\cite{Cheung:2016prv} has been employed to identify the KLA of the maximally-helicity-violating (MHV) sector. Beyond the MHV sector, tensor currents and fusion rules elucidate the KLA of Yang--Mills (YM) theory~\cite{Chen:2019ywi,Chen:2021chy,Brandhuber:2021bsf}. A closed form expression for tree-level CK-dual numerators was obtained from a covariant CK-duality~\cite{Cheung:2021zvb} that identified an underlying kinematic Lorentz algebra. Moreover, it has been shown that the tree-level currents of super Yang--Mills theory come with a KLA~\cite{Ben-Shahar:2021doh}.
    
    Instead of working at the level of amplitudes, we consider CK-duality and the KLA directly at the level of actions~\cite{Bern:2010yg,Borsten:2020zgj,Borsten:2021hua,Borsten:2021rmh}: given tree-level CK-duality, one can always render the action CK-dual using infinitely many auxiliary fields, but at the cost of unitarity, which is broken by Jacobians arising from field redefinitions~\cite{Borsten:2020zgj,Borsten:2021hua,Borsten:2021rmh}. It therefore remains to identify an organizational principle for the resulting tower of auxiliaries and avoid nonlocal field redefinitions altogether. Following~\cite{Reiterer:2019dys}, we find this organizational principle in the form of $\BVbox$-algebras, which ensure the existence of a kinematic Lie algebra.\footnote{A more detailed algebraic characterization of $\BVbox$-algebras, which also closes the gap to CK-duality, is found in~\cite{Borsten:2023ned}.} 
    
    The prime example of a theory with $\BVbox$-algebra is Chern--Simons (CS) theory, and many field theories can be equivalently formulated as CS-type theories on twistor spaces. Using this picture, we are able to reproduce and generalize~e.g.~the results of~\cite{Ben-Shahar:2021zww,Borsten:2023ned,Monteiro:2011pc}. We show that the currents of such field theories come with a KLA which we can readily identify. In many cases, this KLA extends to the amplitudes, and for a special class, it implies conventional CK-duality. For theories for which the anomalies discussed in~\cite{Costello:2021bah} are absent, our arguments extend to the loop level. 
    
    Our results significantly improve the understanding of KLAs and comprise concrete and new examples. They highlight the power of the action perspective on CK-duality as an organizational principle, and our improved algebraic understanding has the power to streamline the computation of the kinematic numerators important in the double copy construction of gravity scattering amplitudes, cf.~\cite{Borsten:2023ned}.
    
    \section{Chern--Simons theory}
    
    We start with the illustrative example of ordinary nonabelian CS theory, which demonstrates all essential features.
    For our purposes, it is convenient to work with Batalin--Vilkovisky (BV) quantization~\cite{Batalin:1981jr,Schwarz:1992nx}, which introduces unphysical fields called antifields in addition to the physical fields and ghosts, and use differential form notation to hide Lorentz indices.
    The BV action of CS theory reads as
    \begin{equation}
		\begin{aligned}
            S_\text{CS}=\int\tr\Big(&\tfrac12 A\wedge \rmd A+\tfrac1{3!}A\wedge [A,A]
            \\
            &+A^+\wedge(\rmd c+[A,c])+\tfrac12c^+\wedge[c,c]\Big),
		\end{aligned}
	\end{equation}
    where $A$ is a gauge potential 1-form; $c$ is the ghost field, a Grassmann-odd scalar function; $A^+$ and $c^+$ are the corresponding antifields (an odd 2-form and an even 3-form); and all fields take values in a color Lie algebra $\frg$. This action is the Maurer--Cartan action for the differential graded (dg) Lie algebra of differential forms with values in $\frg$, $(\Omega^\bullet\otimes \frg,\rmd)$, whose Lie bracket is the wedge product composed with the Lie bracket of $\frg$, whose differential is the exterior derivative $\rmd$, and whose grading is such that $\Omega^p$ carries ghost number $1-p$, see e.g.~\cite{Jurco:2018sby}. After color-stripping, we are left with the dg commutative algebra $(\Omega^p,\rmd)$ of ordinary differential forms under wedge product and exterior derivative.
    
    It is well-known that CS scattering amplitudes are trivial. Instead,~\cite{Ben-Shahar:2021zww} considers correlators of harmonic differential forms. To compute these, note that
    \begin{equation}\label{eq:HodgeLaplacian}
        \rmd\rmd^\dagger+\rmd^\dagger\rmd=-\wave,
    \end{equation}
    where the codifferential $\rmd^\dagger$ is defined as $\rmd^\dagger\alpha=-(-1)^p{\star\rmd\star}\alpha$ for a $p$-form $\alpha$ using the Hodge operator with respect to the Minkowski metric, and $\wave$ the d'Alembertian. The propagator is now given by $\frac{-\rmd^\dagger}{\wave}$. Using~\eqref{eq:HodgeLaplacian}, we may decompose the identity operator on differential forms as
    \begin{equation}\label{eq:contracting_homotopy_3}
        1=\rmd\frac{-\rmd^\dagger}{\wave}+\frac{-\rmd^\dagger}{\wave}\rmd+\Pi_\text{Harm},
    \end{equation}
    where $\Pi_\text{Harm}$ projects onto the harmonic forms. The operator $-\rmd^\dagger$ allows us to introduce a ``derived bracket'' on $\Omega^\bullet$
    \begin{equation}\label{eq:CSGerstenhaber}
        (-1)^p[\alpha,\beta]=-\rmd^\dagger(\alpha\wedge\beta)+\rmd^\dagger\alpha\wedge\beta+(-1)^p\alpha\wedge\rmd^\dagger\beta
    \end{equation}
    for all $\alpha\in\Omega^p$ and $\beta\in\Omega^\bullet$. Since $-\rmd^\dagger$ is a second-order differential operator, $[-,-]$ defines a so-called Gerstenhaber bracket on $\Omega^\bullet$, which is a degree-shifted Lie algebra. Furthermore, the bracket $[-,-]$ maps pairs of physical fields to physical fields and encodes their interactions. Thus, this defines the KLA\footnote{$[1]$ indicates the degree-shift to form an ordinary Lie algebra.}
    \begin{equation}
        \frK=(\Omega^\bullet[1],[-,-])
	\end{equation}
    for correlators of harmonic forms, which therefore can be brought into the form~\eqref{eq:YM_amplitudes_parameterization}. One can show~\cite{koszul1985crochet,Coll:2003ym} that this KLA is isomorphic to the Schouten--Nijenhuis algebra of totally antisymmetric tensor fields, the natural Gerstenhaber algebra on three-dimensional Minkowski space. Truncating $\frK$ to degree~$0$ yields the KLA $\frK_0$ commonly discussed in the literature, which here is the spacetime diffeomorphism algebra.

    The above straightforwardly extends to holomorphic CS theory on $\IC^3$, with the real $p$-forms and the KLA replaced by the complex $(0,p)$-forms and the evident holomorphic version of the Schouten--Nijenhuis algebra, respectively.
    
    More generally, the structure $(\Omega^\bullet,\rmd,\wedge,-\rmd^\dagger)$ is an instance of what is known as a $\BVbox$-algebra~\cite{Reiterer:2019dys,Borsten:2023ned} with $\BBox=\wave$, which we explore below.

    \section{Color--kinematics duality algebraically}
    
    Consider a field theory whose tree amplitudes~\eqref{eq:YM_amplitudes_parameterization} arise from the Feynman diagram expansion of a BV action. The corresponding vector space of fields $\frL$ is graded by the ghost number and so, we may write $\frL=\bigoplus_{p\in\IZ}\frL_p$, where the elements of $\frL_p$ carry ghost number $1-p$. The free part of the action is captured by a kinematic operator, which is a linear map $\sfd$ on $\frL$ with $\sfd^2=0$ that decreases ghost number by $1$ (mapping antifields to fields and fields to zero). All interaction terms are cubic which are captured by a product on $\frL$ that conserves ghost number. Gauge invariance implies that this $\frL$ forms a dg Lie algebra~\cite{Jurco:2018sby,Jurco:2019bvp,Borsten:2021hua}, which generalizes our previous $\Omega^p\otimes \frg$.
    
    The Feynman expansion~\eqref{eq:YM_amplitudes_parameterization} implies that a propagator $\sfh$ exists that inverts $\sfd$ on propagating (off-shell) fields. Thus, the identity operator on fields decomposes as
    \begin{equation}\label{eq:contracting_homotopy}
        1=\sfd\sfh+\sfh\sfd+\Pi_\text{on-shell},
    \end{equation}
    where $\Pi_\text{on-shell}$ is the projector on on-shell fields, generalizing \eqref{eq:contracting_homotopy_3}. It is always possible to choose $\sfh$ such that $\sfh^2=0$~\cite{Markl:0002130}. Splitting $\sfh=\frac{\sfb}{\BBox}$ into the denominator $\BBox$ and numerator $\sfb$ leads to $\sfd\sfb+\sfb\sfd=\BBox$.
    
    Color-stripping now amounts to factorizing $\frL=\frg\otimes\frB$ into the color Lie algebra $\frg$ and a dg commutative algebra  $(\frB,\sfd,\sfm)$ with differential $\sfd$ and product $\sfm$~\cite{Borsten:2021hua}. Denoting the color-stripped propagator also by $\frac{\sfb}{\BBox}$, we have
    \begin{equation}
        \sfd\sfb+\sfb\sfd=\BBox
    \end{equation}
    which generalizes~\eqref{eq:HodgeLaplacian}, and $\BBox$ is a second-order differential operator with respect to $\sfm$. If also $\sfb$ is a second-order differential operator that squares to zero,\footnote{This is true in physically relevant cases~\cite{Borsten:2023ned} and algebraically natural~\cite{Getzler:1994yd}.} the derived bracket defined by
    \begin{equation}\label{eq:kinematic_bracket}
        (-1)^{|x|}[x,y]=\sfb\sfm(x,y)-\sfm(\sfb x,y)-(-1)^{|x|}\sfm(x,\sfb y)
    \end{equation}
    for all $x,y\in\frB$ is a Gerstenhaber bracket on $\frB$ of which~\eqref{eq:CSGerstenhaber} is a special instance. As in the CS case, the KLA (with all BV fields) is then simply
    \begin{equation}
        \frK=(\frB[1],[-,-]).
	\end{equation}
    Truncating $\frK$ to degree~$0$ yields the usual KLA $\frK_0$.

    Mathematically, $(\frB,\sfd,\sfm,\sfb)$ is a BV algebra~\cite{Getzler:1994yd} with Gerstenhaber bracket given by~\eqref{eq:kinematic_bracket}, and $\BBox$ promotes this BV algebra to a $\BVbox$-algebra~\cite{Reiterer:2019dys,Borsten:2023ned}. 
    
    Conversely, in a theory with a $\BVbox$-algebra, the cubic vertices in Feynman diagrams are governed by a KLA and a ``color'' Lie algebra. When $\BBox$ coincides with the d'Alembertian $\wave$, this implies CK-duality (up to potential anomalies). Otherwise, complications may arise such that it is not possible to write the amplitudes in the form~\eqref{eq:YM_amplitudes_parameterization}. Then, we merely speak of a \emph{theory with a KLA}.
    
    Apart from the CS theories already discussed, these ideas extend to field theories whose linearized equations of motions are encoded in a differential with a natural codifferential. This is the case for all theories whose solutions can be described in terms of flat connections on twistor spaces. In the following, we discuss two examples in detail. We stress that even in the absence of an action principle, we still have a $\BVbox$-algebra and, thus, a KLA for the numerators of the corresponding tree-level currents.
    
    \section{Self-dual Yang--Mills theory}
    
    The twistor space $Z$ of maximally supersymmetric (MS) SDYM theory is the superspace $\IR^{4|8}\times\IC P^1$ with $\IC P^1$ the Riemann sphere; see e.g.~\cite{Popov:2004rb,Wolf:2010av} for reviews. Holomorphic CS theory on $Z$ is semi-classically equivalent, see \cite{Mason:2005zm,Boels:2006ir} and~\cite{Wolf:2010av} for details, to MSSDYM theory given by the Siegel action~\cite{Siegel:1992xp} on $\IR^4$,
    \begin{equation}\label{eq:action_Siegel}
        S_\text{Siegel}=\int  \tr \Big(G^-\wedge F-\tfrac12\phi\wave\phi+\chi\nabla\chi+\tfrac12\phi[\chi,\chi]\Big),
	\end{equation}
    where $F$ is the gluon field strength 2-form, the Lagrange multiplier $G^-$ is a antiself-dual 2-form, $\phi$ denotes the six scalar fields, and $\chi$ denotes the four gluinos; all fields transform under the adjoint representation of the gauge group. Adopting coordinates $(x^{\alpha\dot\alpha},\eta_i^{\dot\alpha})$ on $\IR^{4|8}$, with $x^{\alpha\dot\alpha}$ a commuting 4-vector and $\eta^{\dot\alpha}_i$ anticommuting spinors, and homogeneous coordinates $\lambda_{\dot\alpha}$ on $\IC P^1$ ($\alpha,\dot\alpha=1,2$; $i=1,\dotsc,4$), the antiholomorphic vector fields on $Z$ are spanned by
    \begin{equation}\label{eq:vector_fields}
        (\hat E_\alpha,\hat E^i,\hat E_0)=\left(\lambda^{\dot\alpha}\parder{x^{\alpha\dot \alpha}},\lambda^{\dot\alpha}\parder{\eta^{\dot\alpha}_i},|\lambda|^2\lambda_{\dot \beta}\parder{\hat\lambda_{\dot\beta}}\right)\!,
    \end{equation}
    where $|\lambda|^2=\lambda_{\dot\alpha}\hat\lambda^{\dot\alpha}$ (with $\hat\lambda^{\dot\alpha}=\bar\lambda_{\dot\alpha}$). The holomorphic vector fields $(E_\alpha,E^i,E_0)$ are the corresponding conjugates. We denote the antiholomorphic differential 1-forms dual to~\eqref{eq:vector_fields} by $(\hat e^\alpha,\hat e_i,\hat e^0)$. We then have the BV action
    \begin{equation}\label{eq:action_hCS1}
		\begin{aligned}
            S_\text{hCS}&=\int\Omega\wedge\tr\Big(\tfrac12 A\wedge \bar\dpar_{\rm red}A+\tfrac1{3!}A\wedge [A,A]
            \\
            &\hspace{1.5cm}+A^+\wedge(\bar\dpar_{\rm red}c+[A,c])+\tfrac12c^+\wedge[c,c]\Big),
		\end{aligned}
	\end{equation}
    where $\Omega$ is the holomorphic volume form on twistor space~\cite{Witten:2003nn} and $\bar\dpar_{\rm red}=\hat e^\alpha\hat E_\alpha+\hat e^0\hat E_0$ is the Dolbeault differential restricted to legs along commuting coordinates. The fields are $\frg$-valued antiholomorphic differential forms on $Z$ that are holomorphic with respect to the anticommuting coordinates $\eta_i=\eta^{\dot\alpha}_i\lambda_{\dot\alpha}$ and have no legs along antiholomorphic anticommuting coordinates. 
    
    The space of color-stripped fields forms a dg commutative algebra, with product given by the wedge product and differential $\bar\dpar_{\rm red}$. It further forms a $\BVbox$-algebra $\frB_\text{SDYM}$ together with the differential operator
    \begin{equation}
        \sfb=-\frac{4}{|\lambda|^2}\varepsilon^{\alpha\beta}\iota_{E_\alpha}\iota_{\hat E_\beta}\partial_{\rm red}+2\varepsilon^{\alpha\beta}\iota_{\hat E_\alpha}\iota_{\hat E_\beta}\hat e^0\wedge,
    \end{equation}
    where $\iota_X$ denotes the contraction of a differential form with a vector field $X$. A quick calculation shows that $\bar\dpar_{\rm red}\sfb+\sfb\bar\dpar_{\rm red}=\wave{}_{\IR^4}$.
    
    Recall that the actions~\eqref{eq:action_hCS1} and~\eqref{eq:action_Siegel} are equivalent, i.e.~they share the same tree-level amplitudes. We can compute these by embedding external states on $\IR^{4}$, given by harmonic gauge potentials, into $A\in\Omega^{0|0,1|0}_{\rm red}$, respecting the gauge condition $\sfb A=0$. We then use the trivial Feynman rules derived from~\eqref{eq:action_hCS1} together with the propagator $\sfh=\frac{\sfb}{\wave{}_{\IR^4}}$. This Feynman diagram expansion manifests the KLA contained in $\frB_\text{SDYM}$ with $\BBox=\wave{}_{\IR^4}$. Hence, MSSDYM theory possesses CK-duality, and the twistor action produces a CK-duality-manifesting spacetime action for MSSDYM theory after Kaluza--Klein (KK)-expanding along $\IC P^1$. Integrating out the KK tower of auxiliary fields reproduces the Siegel action.
    
    The full KLA $\frK$ is isomorphic to the Schouten--Nijenhuis-type Lie algebra $\tilde\frK$ of bosonic holomorphic totally antisymmetric tensor fields on $Z$ with Lie bracket $[U,V]_{\rm red}=(U^\alpha E_\alpha V^\beta-V^\alpha E_\alpha U^\beta)E_\beta$.
    
    This construction generalizes to dimensionally reduced SDYM theory and theories with any amount of supersymmetry, following~\cite{Saemann:2004tt,Popov:2004nk} or, e.g.~\cite{Popov:2005uv}.
    
    To match the literature, note that the lowest order in $\eta_i$ of the gauge potential $A$ describes the gluon. Since to this order, $\iota_{\hat E_0}A$ can be gauged away~\cite[\S5.2]{Wolf:2010av}, only holomorphic multivector fields spanned by the $E_\alpha$ in the Schouten--Nijenhuis-type Lie algebra $\tilde\frK$ contribute to $\tilde\frK_0$. On spacetime, these parameterize translations along self-dual planes spanned by $(\parder u,\parder w)=(\parder{x^{1\dot 1}},\parder{x^{2\dot 1}})$, reproducing the KLA identified in~\cite{Monteiro:2011pc}.
    
    Beyond trees, (MS)SDYM theory possesses finite 1-loop amplitudes. Unlike SDYM theory, the twistorial action for MSSDYM theory captures the correct 1-loop amplitudes~\cite{Costello:2021bah}. Our arguments therefore also demonstrate CK-duality for MSSDYM theory 1-loop amplitude integrands.
    
    \section{Maximally-supersymmetric Yang--Mills theory}
    
    Similar arguments also hold for full maximally-supersymmetric Yang--Mills (MSYM) theory.\footnote{Recall that $\caN=3$ SYM theory is perturbatively equivalent to $\caN=4$ SYM theory.} However, the KLA will be such that CK-duality is not immediate. 

    In this case, the twistor space is a CR-manifold (a generalization of the notion of a complex manifold, cf.~\cite{Boggess:1991aa}), namely the CR ambitwistor space $L\cong\IR^{4|24}\times\IC P^1\times\IC P^1$. Holomorphic CS theory on $L$ is semi-classically equivalent to four-dimensional MSYM theory~\cite{Movshev:2004ub,Mason:2005kn}. We use Cartesian coordinates in spinor notation $(x^{\alpha\dot\alpha},\eta_i^{\dot\alpha},\theta^{i\alpha})$ on $\IR^{4|24}$ and homogeneous coordinates $(\lambda_{\dot\alpha},\mu_\alpha)$ on $\IC P^1\times\IC P^1$. The antiholomorphic vector fields
    \begin{equation}
        (\hat E_\rmF,\hat E_\rmL,\hat E_\rmR)=\left(\mu^\alpha \lambda^{\dot \alpha}\parder{x^{\alpha\dot \alpha}},|\lambda|^2\lambda_{\dot \beta}\parder{\hat\lambda_{\dot\beta}},|\mu|^2\mu_{\beta}\parder{\hat\mu_{\beta}}\right)
    \end{equation}
    form a basis of the space of antiholomorphic vector fields along commuting directions; we also define the conjugate holomorphic vector fields $(E_\rmF,E_\rmL,E_\rmR)$ and dual antiholomorphic 1-forms $(\hat e^\rmF,\hat e^\rmL,\hat e^\rmR)$.
    
    The relevant action here is again of the form~\eqref{eq:action_hCS1}, with $Z$ replaced by $L$, $\Omega^{3|4,0|0}$ replaced by the holomorphic measure identified in~\cite{Movshev:2004ub,Mason:2005kn}, and the dg commutative algebra $(\Omega^{0|0,\bullet|0}_{\rm red},\bar\dpar_{\rm red})$ replaced by the restricted bosonic CR differential forms $\Omega^{0|0,\bullet|0}_\text{CR}$ that depend holomorphically on $(\eta_i,\theta^i)=(\eta_i^{\dot\alpha}\lambda_{\dot\alpha},\theta^{i\alpha}\mu_\alpha)$ for $i=1,2,3$ and with no antiholomorphic fermionic directions, which are endowed with the differential $\bar\dpar_\text{CR}=\hat e^\rmF\hat E_\rmF+\hat e^\rmL\hat E_\rmL+\hat e^\rmR\hat E_\rmR$. The second-order differential operator
    \begin{equation}
        \sfb=-\frac{8}{|\lambda|^2|\mu|^2}\iota_{E_\rmF}\iota_{{\hat E}_\rmF}\dpar_\text{CR}
    \end{equation}
    leads to
    \begin{equation}
        \begin{aligned}
            \BBox&=\bar\dpar_\text{CR}\sfb+\sfb\bar\dpar_\text{CR}
            \\
            &\,=\wave_{\IR^4}+8\frac{\mu^{(\alpha}\hat\mu^{\beta)}\lambda^{(\dot\alpha}\hat\lambda^{\dot\beta)}}{|\lambda|^2|\mu|^2}\parder{x^{\alpha\dot\alpha}}\parder{x^{\beta\dot\beta}}.
        \end{aligned}
    \end{equation}
    Thus, we obtain the $\BVbox$-algebra $\frB_\text{SYM}=(\Omega^{0|0,\bullet|0}_\text{CR},$ $\bar \dpar_\text{CR},\wedge,\sfb)$ containing a KLA of evident Cauchy--Riemann automorphisms of $L$.
    
    While $L$ is not compatible with Wick rotation, $\frB_\text{SYM}$ and the contained KLA are: KK-expand the theory along $\IC P^1\times \IC P^1$, obtaining a cubic field theory with an infinite tower of KK fields on $\IR^4$, on which $(\bar\dpar_\text{CR},\sfb,\BBox)$ act as ``$(\infty\times\infty)$-matrices of differential operators.''
    
    The complexified KK fields all carry complex $\sfSpin(4)$-representations\footnote{Unlike ordinary KK-expansions, compact directions here carry nontrivial Lorentz representations: Wick rotation destroys the geometric interpretation.}, and imposing reality conditions suitable for Minkowski space Wick-rotates both the fields and the operators $(\bar\dpar_\text{CR},\sfb,\BBox)$.\footnote{We note that KK expansion/Wick rotation preserve semi-classical equivalence between Cauchy--Riemann CS theory and MSYM theory. Perturbation theory with propagator $\frac{\sfb}{\BBox}$ and gauge $\sfb A=0$ for the field $A$ reproduces MSYM amplitudes on four-dimensional Minkowski space: semi-classical equivalence fixes the interaction vertices; the propagator $\frac{\sfb}{\BBox}$ is the inverse of the kinematic operator almost-everywhere (i.e.~modulo measure-zero sets) in momentum space.} 
    
    In view of CK-duality, our above propagator involving the inverse of $\BBox$, 
    \begin{equation}
        \frac1\BBox=\frac{\eta^{MN}-K^{MN}_{\mu\nu}\frac{k^\mu k^\nu}{k^2}+K^{MP}_{\mu\nu}K_P{}^{N}{}_{\rho\sigma}\frac{k^\mu k^\nu k^\rho k^\sigma}{k^4}-\dotsb}{k^2},
    \end{equation}
    where $M,N,\ldots$ label KK modes, has a striking similarity to the YM propagator $\sfh^{\mu\nu}_\xi=\frac{1}{k^2}\eta^{\mu\nu}+(1-\xi)\frac{k^\mu k^\nu}{k^4}$ in $R_\xi$-gauge in that both lead to unphysical singularities (e.g.~$\frac{k^\mu k^\nu}{k^4}$ for $\sfh^{\mu\nu}_\xi$) in individual Feynman diagrams. These singularities signal the propagation of unphysical longitudinal modes. With all external states physical, their contributions have to cancel, and for $R_\xi$-gauge this follows from Ward identities. It is natural to expect that the same occurs for (potentially Wick-rotated) Cauchy--Riemann CS theory, and there is a KK tower of generalized Ward identities that allows to replace the propagator $\frac{\sfb}{\BBox}$ with $\frac{\sfb}{\square}$. If true, then the $\BVbox$-algebra guarantees CK-dual numerators: the numerators computed with the propagator $\frac{\sfb}{\square}$ are automatically CK-dual. We study this in upcoming work.
    
    The extension to the loop level depends on the assumption that the relevant (ambi)twistor theories correctly describe the loop amplitudes. The semi-classical equivalences between spacetime field theories and twistorial CS-type theories only extends to the loop level if certain twistor space anomalies vanish~\cite{Costello:2021bah}. Provided that Cauchy--Riemann CS theory on $L$ is anomaly-free with no further problems reducing the path-integral measure from twistor space fields to spacetime fields, then Cauchy--Riemann CS theory on $L$ captures also loop amplitudes, and we obtain a loop-level KLA.
    
    Let us sketch an argument that this is true for the Cauchy--Riemann CS anomaly. A codimension~$k$ Levi-flat Cauchy--Riemann manifold $M$ foliates into holomorphic leaves $M_t$. If the space of leaves $T$ is a $k$-dimensional manifold, one easily checks that $L$ satisfies these conditions, then the Cauchy--Riemann CS partition function is $\log Z(M)=\int_{t\in T}\omega\log Z(M_t)$, where $\omega$ is a volume form (defining the path integral measure) on $T$, and where $Z(M_t)$ is the partition function of holomorphic CS theory (with the same field content) on $M_t$: the full theory on $M$ is anomaly-free if the corresponding holomorphic theory on $M_t$ is anomaly-free.
    
    Thus, it suffices to study holomorphic CS theory on $M_t$, or even (as anomalies are integrals of local objects)  on small patch in $M_t$, where global issues (e.g.~nonzero-degree bundles) disappear. In the weak-coupling limit, anomaly contributions from bosons and fermions are equal and opposite~\cite{Costello:2021bah}, since their linearized actions in the presence of an external gauge field coincide up to statistics: supersymmetry ensures anomaly cancellation in holomorphic CS theory and Cauchy--Riemann CS theory. Without supersymmetry, this argument fails.
    
    Indeed, for nonsupersymmetric twistorial holomorphic CS theory (semi-classically equivalent to SDYM theory) there is an anomaly~\cite{Costello:2021bah} that, via the preceding argument, implies  an anomaly for nonsupersymmetric twistorial Cauchy--Riemann CS theory (semi-classically equivalent to YM theory). Even if the KLA algebra \emph{does} imply tree-level CK-duality, it will be anomalous. This is consistent with, and elucidates, the conclusion that CK-duality can be realized as an anomalous  symmetry of a semi-classically equivalent YM--BV action~\cite{Borsten:2021rmh}, as well as the proof (by exhaustion) that there are no CK-dual four-point two-loop numerators for bosonic YM theory assuming that they can be derived from local Feynman rules~\cite{Bern:2015ooa}. 

    \
    
    \noindent
    \textbf{Data Management.}
    No additional research data beyond the data presented and cited in this work are needed to validate the research findings in this work. For the purpose of open access, the authors have applied a Creative Commons Attribution (CC-BY) license to any Author Accepted Manuscript version arising.

    \
    
    \begin{acknowledgments}
        \noindent
        \textbf{Acknowledgments.}
        H.K. and C.S.~were supported by the Leverhulme Research Project Grant RPG-2018-329. B.J.~was supported by the GA\v{C}R Grant EXPRO 19-28268X.
    \end{acknowledgments}

    \bibliography{bigone}

\begin{thebibliography}{64}%
\makeatletter
\providecommand \@ifxundefined [1]{%
 \@ifx{#1\undefined}
}%
\providecommand \@ifnum [1]{%
 \ifnum #1\expandafter \@firstoftwo
 \else \expandafter \@secondoftwo
 \fi
}%
\providecommand \@ifx [1]{%
 \ifx #1\expandafter \@firstoftwo
 \else \expandafter \@secondoftwo
 \fi
}%
\providecommand \natexlab [1]{#1}%
\providecommand \enquote  [1]{``#1''}%
\providecommand \bibnamefont  [1]{#1}%
\providecommand \bibfnamefont [1]{#1}%
\providecommand \citenamefont [1]{#1}%
\providecommand \href@noop [0]{\@secondoftwo}%
\providecommand \href [0]{\begingroup \@sanitize@url \@href}%
\providecommand \@href[1]{\@@startlink{#1}\@@href}%
\providecommand \@@href[1]{\endgroup#1\@@endlink}%
\providecommand \@sanitize@url [0]{\catcode `\\12\catcode `\$12\catcode
  `\&12\catcode `\#12\catcode `\^12\catcode `\_12\catcode `\%12\relax}%
\providecommand \@@startlink[1]{}%
\providecommand \@@endlink[0]{}%
\providecommand \url  [0]{\begingroup\@sanitize@url \@url }%
\providecommand \@url [1]{\endgroup\@href {#1}{\urlprefix }}%
\providecommand \urlprefix  [0]{URL }%
\providecommand \Eprint [0]{\href }%
\providecommand \doibase [0]{https://doi.org/}%
\providecommand \selectlanguage [0]{\@gobble}%
\providecommand \bibinfo  [0]{\@secondoftwo}%
\providecommand \bibfield  [0]{\@secondoftwo}%
\providecommand \translation [1]{[#1]}%
\providecommand \BibitemOpen [0]{}%
\providecommand \bibitemStop [0]{}%
\providecommand \bibitemNoStop [0]{.\EOS\space}%
\providecommand \EOS [0]{\spacefactor3000\relax}%
\providecommand \BibitemShut  [1]{\csname bibitem#1\endcsname}%
\let\auto@bib@innerbib\@empty
\bibitem [{\citenamefont {Reiterer}(2019)}]{Reiterer:2019dys}%
  \BibitemOpen
  \bibfield  {author} {\bibinfo {author} {\bibfnamefont {M.}~\bibnamefont
  {Reiterer}},\ }\bibfield  {title} {\bibinfo {title} {A homotopy {BV} algebra
  for {Y}ang--{M}ills and color--kinematics},\ }\href@noop {} {\  (\bibinfo
  {year} {2019})},\ \Eprint {https://arxiv.org/abs/1912.03110}
  {arXiv:1912.03110 [math-ph]} \BibitemShut {NoStop}%
\bibitem [{\citenamefont {Bern}\ \emph {et~al.}(2008)\citenamefont {Bern},
  \citenamefont {Carrasco},\ and\ \citenamefont {Johansson}}]{Bern:2008qj}%
  \BibitemOpen
  \bibfield  {author} {\bibinfo {author} {\bibfnamefont {Z.}~\bibnamefont
  {Bern}}, \bibinfo {author} {\bibfnamefont {J.~J.~M.}\ \bibnamefont
  {Carrasco}},\ and\ \bibinfo {author} {\bibfnamefont {H.}~\bibnamefont
  {Johansson}},\ }\bibfield  {title} {\bibinfo {title} {New relations for
  gauge-theory amplitudes},\ }\href
  {https://doi.org/10.1103/PhysRevD.78.085011} {\bibfield  {journal} {\bibinfo
  {journal} {Phys. Rev. D}\ }\textbf {\bibinfo {volume} {78}},\ \bibinfo
  {pages} {085011} (\bibinfo {year} {2008})},\ \Eprint
  {https://arxiv.org/abs/0805.3993} {arXiv:0805.3993 [hep-ph]} \BibitemShut
  {NoStop}%
\bibitem [{\citenamefont {Bern}\ \emph
  {et~al.}(2010{\natexlab{a}})\citenamefont {Bern}, \citenamefont {Carrasco},\
  and\ \citenamefont {Johansson}}]{Bern:2010ue}%
  \BibitemOpen
  \bibfield  {author} {\bibinfo {author} {\bibfnamefont {Z.}~\bibnamefont
  {Bern}}, \bibinfo {author} {\bibfnamefont {J.~J.~M.}\ \bibnamefont
  {Carrasco}},\ and\ \bibinfo {author} {\bibfnamefont {H.}~\bibnamefont
  {Johansson}},\ }\bibfield  {title} {\bibinfo {title} {Perturbative quantum
  gravity as a double copy of gauge theory},\ }\href
  {https://doi.org/10.1103/PhysRevLett.105.061602} {\bibfield  {journal}
  {\bibinfo  {journal} {Phys. Rev. Lett.}\ }\textbf {\bibinfo {volume} {105}},\
  \bibinfo {pages} {061602} (\bibinfo {year} {2010}{\natexlab{a}})},\ \Eprint
  {https://arxiv.org/abs/1004.0476} {arXiv:1004.0476 [hep-th]} \BibitemShut
  {NoStop}%
\bibitem [{\citenamefont {Carrasco}()}]{Carrasco:2015iwa}%
  \BibitemOpen
  \bibfield  {author} {\bibinfo {author} {\bibfnamefont {J.~J.~M.}\
  \bibnamefont {Carrasco}},\ }\bibfield  {title} {\bibinfo {title} {Gauge and
  gravity amplitude relations},\ }\href
  {https://doi.org/10.1142/9789814678766_0011} {\ ,\ \bibinfo {pages}
  {477}}\bibinfo {note} {In: ``Theoretical Advanced Study Institute in
  Elementary Particle Physics: Journeys Through the Precision Frontier:
  Amplitudes for Colliders''},\ \Eprint {https://arxiv.org/abs/1506.00974}
  {arXiv:1506.00974 [hep-th]} \BibitemShut {NoStop}%
\bibitem [{\citenamefont {Borsten}(2020)}]{Borsten:2020bgv}%
  \BibitemOpen
  \bibfield  {author} {\bibinfo {author} {\bibfnamefont {L.}~\bibnamefont
  {Borsten}},\ }\bibfield  {title} {\bibinfo {title} {Gravity as the square of
  gauge theory: a review},\ }\href {https://doi.org/10.1007/s40766-020-00003-6}
  {\bibfield  {journal} {\bibinfo  {journal} {Riv. Nuovo Cim.}\ }\textbf
  {\bibinfo {volume} {43}},\ \bibinfo {pages} {97} (\bibinfo {year}
  {2020})}\BibitemShut {NoStop}%
\bibitem [{\citenamefont {Bern}\ \emph
  {et~al.}(2019{\natexlab{a}})\citenamefont {Bern}, \citenamefont {Carrasco},
  \citenamefont {Chiodaroli}, \citenamefont {Johansson},\ and\ \citenamefont
  {Roiban}}]{Bern:2019prr}%
  \BibitemOpen
  \bibfield  {author} {\bibinfo {author} {\bibfnamefont {Z.}~\bibnamefont
  {Bern}}, \bibinfo {author} {\bibfnamefont {J.~J.}\ \bibnamefont {Carrasco}},
  \bibinfo {author} {\bibfnamefont {M.}~\bibnamefont {Chiodaroli}}, \bibinfo
  {author} {\bibfnamefont {H.}~\bibnamefont {Johansson}},\ and\ \bibinfo
  {author} {\bibfnamefont {R.}~\bibnamefont {Roiban}},\ }\bibfield  {title}
  {\bibinfo {title} {The duality between color and kinematics and its
  applications},\ }\href@noop {} {\  (\bibinfo {year} {2019}{\natexlab{a}})},\
  \Eprint {https://arxiv.org/abs/1909.01358} {arXiv:1909.01358 [hep-th]}
  \BibitemShut {NoStop}%
\bibitem [{\citenamefont {Adamo}\ \emph {et~al.}(2022)\citenamefont {Adamo},
  \citenamefont {Carrasco}, \citenamefont {Carrillo-Gonz\'{a}lez},
  \citenamefont {Chiodaroli}, \citenamefont {Elvang}, \citenamefont
  {Johansson}, \citenamefont {O'Connell}, \citenamefont {Roiban},\ and\
  \citenamefont {Schlotterer}}]{Adamo:2022dcm}%
  \BibitemOpen
  \bibfield  {author} {\bibinfo {author} {\bibfnamefont {T.}~\bibnamefont
  {Adamo}}, \bibinfo {author} {\bibfnamefont {J.~J.~M.}\ \bibnamefont
  {Carrasco}}, \bibinfo {author} {\bibfnamefont {M.}~\bibnamefont
  {Carrillo-Gonz\'{a}lez}}, \bibinfo {author} {\bibfnamefont {M.}~\bibnamefont
  {Chiodaroli}}, \bibinfo {author} {\bibfnamefont {H.}~\bibnamefont {Elvang}},
  \bibinfo {author} {\bibfnamefont {H.}~\bibnamefont {Johansson}}, \bibinfo
  {author} {\bibfnamefont {D.}~\bibnamefont {O'Connell}}, \bibinfo {author}
  {\bibfnamefont {R.}~\bibnamefont {Roiban}},\ and\ \bibinfo {author}
  {\bibfnamefont {O.}~\bibnamefont {Schlotterer}},\ }\bibfield  {title}
  {\bibinfo {title} {Snowmass white paper: The double copy and its
  applications},\ }\bibfield  {booktitle} {\emph {\bibinfo {booktitle} {{2022
  Snowmass Summer Study}}},\ }\href@noop {} {\  (\bibinfo {year} {2022})},\
  \Eprint {https://arxiv.org/abs/2204.06547} {arXiv:2204.06547 [hep-th]}
  \BibitemShut {NoStop}%
\bibitem [{\citenamefont {Bern}\ \emph {et~al.}(2022)\citenamefont {Bern},
  \citenamefont {Carrasco}, \citenamefont {Chiodaroli}, \citenamefont
  {Johansson},\ and\ \citenamefont {Roiban}}]{Bern:2022wqg}%
  \BibitemOpen
  \bibfield  {author} {\bibinfo {author} {\bibfnamefont {Z.}~\bibnamefont
  {Bern}}, \bibinfo {author} {\bibfnamefont {J.~J.~M.}\ \bibnamefont
  {Carrasco}}, \bibinfo {author} {\bibfnamefont {M.}~\bibnamefont
  {Chiodaroli}}, \bibinfo {author} {\bibfnamefont {H.}~\bibnamefont
  {Johansson}},\ and\ \bibinfo {author} {\bibfnamefont {R.}~\bibnamefont
  {Roiban}},\ }\bibfield  {title} {\bibinfo {title} {The {SAGEX} review on
  scattering amplitudes, chapter 2: {A}n invitation to color--kinematics
  duality and the double copy},\ }\href@noop {} {\  (\bibinfo {year} {2022})},\
  \Eprint {https://arxiv.org/abs/2203.13013} {arXiv:2203.13013 [hep-th]}
  \BibitemShut {NoStop}%
\bibitem [{\citenamefont {Bern}\ \emph {et~al.}(2021)\citenamefont {Bern},
  \citenamefont {Parra-Martinez}, \citenamefont {Roiban}, \citenamefont {Ruf},
  \citenamefont {Shen}, \citenamefont {Solon},\ and\ \citenamefont
  {Zeng}}]{Bern:2021dqo}%
  \BibitemOpen
  \bibfield  {author} {\bibinfo {author} {\bibfnamefont {Z.}~\bibnamefont
  {Bern}}, \bibinfo {author} {\bibfnamefont {J.}~\bibnamefont
  {Parra-Martinez}}, \bibinfo {author} {\bibfnamefont {R.}~\bibnamefont
  {Roiban}}, \bibinfo {author} {\bibfnamefont {M.~S.}\ \bibnamefont {Ruf}},
  \bibinfo {author} {\bibfnamefont {C.-H.}\ \bibnamefont {Shen}}, \bibinfo
  {author} {\bibfnamefont {M.~P.}\ \bibnamefont {Solon}},\ and\ \bibinfo
  {author} {\bibfnamefont {M.}~\bibnamefont {Zeng}},\ }\bibfield  {title}
  {\bibinfo {title} {Scattering amplitudes and conservative binary dynamics at
  ${\mathcal{o}}(g^4)$},\ }\href
  {https://doi.org/10.1103/PhysRevLett.126.171601} {\bibfield  {journal}
  {\bibinfo  {journal} {Phys. Rev. Lett.}\ }\textbf {\bibinfo {volume} {126}},\
  \bibinfo {pages} {171601} (\bibinfo {year} {2021})},\ \Eprint
  {https://arxiv.org/abs/2101.07254} {arXiv:2101.07254 [hep-th]} \BibitemShut
  {NoStop}%
\bibitem [{\citenamefont {Bern}\ \emph
  {et~al.}(2019{\natexlab{b}})\citenamefont {Bern}, \citenamefont {Cheung},
  \citenamefont {Roiban}, \citenamefont {Shen}, \citenamefont {Solon},\ and\
  \citenamefont {Zeng}}]{Bern:2019crd}%
  \BibitemOpen
  \bibfield  {author} {\bibinfo {author} {\bibfnamefont {Z.}~\bibnamefont
  {Bern}}, \bibinfo {author} {\bibfnamefont {C.}~\bibnamefont {Cheung}},
  \bibinfo {author} {\bibfnamefont {R.}~\bibnamefont {Roiban}}, \bibinfo
  {author} {\bibfnamefont {C.-H.}\ \bibnamefont {Shen}}, \bibinfo {author}
  {\bibfnamefont {M.~P.}\ \bibnamefont {Solon}},\ and\ \bibinfo {author}
  {\bibfnamefont {M.}~\bibnamefont {Zeng}},\ }\bibfield  {title} {\bibinfo
  {title} {Black hole binary dynamics from the double copy and effective
  theory},\ }\href {https://doi.org/10.1007/JHEP10(2019)206} {\bibfield
  {journal} {\bibinfo  {journal} {JHEP}\ }\textbf {\bibinfo {volume} {1910}},\
  \bibinfo {pages} {206}},\ \Eprint {https://arxiv.org/abs/1908.01493}
  {arXiv:1908.01493 [hep-th]} \BibitemShut {NoStop}%
\bibitem [{\citenamefont {Bern}\ \emph
  {et~al.}(2019{\natexlab{c}})\citenamefont {Bern}, \citenamefont {Cheung},
  \citenamefont {Roiban}, \citenamefont {Shen}, \citenamefont {Solon},\ and\
  \citenamefont {Zeng}}]{Bern:2019nnu}%
  \BibitemOpen
  \bibfield  {author} {\bibinfo {author} {\bibfnamefont {Z.}~\bibnamefont
  {Bern}}, \bibinfo {author} {\bibfnamefont {C.}~\bibnamefont {Cheung}},
  \bibinfo {author} {\bibfnamefont {R.}~\bibnamefont {Roiban}}, \bibinfo
  {author} {\bibfnamefont {C.-H.}\ \bibnamefont {Shen}}, \bibinfo {author}
  {\bibfnamefont {M.~P.}\ \bibnamefont {Solon}},\ and\ \bibinfo {author}
  {\bibfnamefont {M.}~\bibnamefont {Zeng}},\ }\bibfield  {title} {\bibinfo
  {title} {Scattering amplitudes and the conservative hamiltonian for binary
  systems at third post-minkowskian order},\ }\href
  {https://doi.org/10.1103/PhysRevLett.122.201603} {\bibfield  {journal}
  {\bibinfo  {journal} {Phys. Rev. Lett.}\ }\textbf {\bibinfo {volume} {122}},\
  \bibinfo {pages} {201603} (\bibinfo {year} {2019}{\natexlab{c}})},\ \Eprint
  {https://arxiv.org/abs/1901.04424} {arXiv:1901.04424 [hep-th]} \BibitemShut
  {NoStop}%
\bibitem [{\citenamefont {Bjerrum-Bohr}\ \emph {et~al.}(2009)\citenamefont
  {Bjerrum-Bohr}, \citenamefont {Damgaard},\ and\ \citenamefont
  {Vanhove}}]{BjerrumBohr:2009rd}%
  \BibitemOpen
  \bibfield  {author} {\bibinfo {author} {\bibfnamefont {N.~E.~J.}\
  \bibnamefont {Bjerrum-Bohr}}, \bibinfo {author} {\bibfnamefont {P.~H.}\
  \bibnamefont {Damgaard}},\ and\ \bibinfo {author} {\bibfnamefont
  {P.}~\bibnamefont {Vanhove}},\ }\bibfield  {title} {\bibinfo {title} {Minimal
  basis for gauge theory amplitudes},\ }\href
  {https://doi.org/10.1103/PhysRevLett.103.161602} {\bibfield  {journal}
  {\bibinfo  {journal} {Phys. Rev. Lett.}\ }\textbf {\bibinfo {volume} {103}},\
  \bibinfo {pages} {161602} (\bibinfo {year} {2009})},\ \Eprint
  {https://arxiv.org/abs/0907.1425} {arXiv:0907.1425 [hep-th]} \BibitemShut
  {NoStop}%
\bibitem [{\citenamefont {Stieberger}(2009)}]{Stieberger:2009hq}%
  \BibitemOpen
  \bibfield  {author} {\bibinfo {author} {\bibfnamefont {S.}~\bibnamefont
  {Stieberger}},\ }\bibfield  {title} {\bibinfo {title} {{Open \& closed
  vs.~pure open string disk amplitudes}},\ }\href@noop {} {\  (\bibinfo {year}
  {2009})},\ \Eprint {https://arxiv.org/abs/0907.2211} {arXiv:0907.2211
  [hep-th]} \BibitemShut {NoStop}%
\bibitem [{\citenamefont {Feng}\ \emph {et~al.}(2011)\citenamefont {Feng},
  \citenamefont {Huang},\ and\ \citenamefont {Jia}}]{Feng:2010my}%
  \BibitemOpen
  \bibfield  {author} {\bibinfo {author} {\bibfnamefont {B.}~\bibnamefont
  {Feng}}, \bibinfo {author} {\bibfnamefont {R.}~\bibnamefont {Huang}},\ and\
  \bibinfo {author} {\bibfnamefont {Y.}~\bibnamefont {Jia}},\ }\bibfield
  {title} {\bibinfo {title} {Gauge amplitude identities by on-shell recursion
  relation in {S}-matrix program},\ }\href
  {https://doi.org/10.1016/j.physletb.2010.11.011} {\bibfield  {journal}
  {\bibinfo  {journal} {Phys. Lett. B}\ }\textbf {\bibinfo {volume} {695}},\
  \bibinfo {pages} {350} (\bibinfo {year} {2011})},\ \Eprint
  {https://arxiv.org/abs/1004.3417} {arXiv:1004.3417 [hep-th]} \BibitemShut
  {NoStop}%
\bibitem [{\citenamefont {Bjerrum-Bohr}\ \emph {et~al.}(2011)\citenamefont
  {Bjerrum-Bohr}, \citenamefont {Damgaard}, \citenamefont {S{\o{}}ndergaard},\
  and\ \citenamefont {Vanhove}}]{BjerrumBohr:2010hn}%
  \BibitemOpen
  \bibfield  {author} {\bibinfo {author} {\bibfnamefont {N.~E.~J.}\
  \bibnamefont {Bjerrum-Bohr}}, \bibinfo {author} {\bibfnamefont {P.~H.}\
  \bibnamefont {Damgaard}}, \bibinfo {author} {\bibfnamefont {T.}~\bibnamefont
  {S{\o{}}ndergaard}},\ and\ \bibinfo {author} {\bibfnamefont {P.}~\bibnamefont
  {Vanhove}},\ }\bibfield  {title} {\bibinfo {title} {The momentum kernel of
  gauge and gravity theories},\ }\href
  {https://doi.org/10.1007/JHEP01(2011)001} {\bibfield  {journal} {\bibinfo
  {journal} {JHEP}\ }\textbf {\bibinfo {volume} {1101}},\ \bibinfo {pages}
  {001}},\ \Eprint {https://arxiv.org/abs/1010.3933} {arXiv:1010.3933 [hep-th]}
  \BibitemShut {NoStop}%
\bibitem [{\citenamefont {Broedel}\ and\ \citenamefont
  {Carrasco}(2011)}]{Broedel:2011pd}%
  \BibitemOpen
  \bibfield  {author} {\bibinfo {author} {\bibfnamefont {J.}~\bibnamefont
  {Broedel}}\ and\ \bibinfo {author} {\bibfnamefont {J.~J.~M.}\ \bibnamefont
  {Carrasco}},\ }\bibfield  {title} {\bibinfo {title} {Virtuous trees at five
  and six points for {Y}ang--{M}ills and gravity},\ }\href
  {https://doi.org/10.1103/PhysRevD.84.085009} {\bibfield  {journal} {\bibinfo
  {journal} {Phys. Rev. D}\ }\textbf {\bibinfo {volume} {84}},\ \bibinfo
  {pages} {085009} (\bibinfo {year} {2011})},\ \Eprint
  {https://arxiv.org/abs/1107.4802} {arXiv:1107.4802 [hep-th]} \BibitemShut
  {NoStop}%
\bibitem [{\citenamefont {Mafra}\ \emph {et~al.}(2011)\citenamefont {Mafra},
  \citenamefont {Schlotterer},\ and\ \citenamefont
  {Stieberger}}]{Mafra:2011kj}%
  \BibitemOpen
  \bibfield  {author} {\bibinfo {author} {\bibfnamefont {C.~R.}\ \bibnamefont
  {Mafra}}, \bibinfo {author} {\bibfnamefont {O.}~\bibnamefont {Schlotterer}},\
  and\ \bibinfo {author} {\bibfnamefont {S.}~\bibnamefont {Stieberger}},\
  }\bibfield  {title} {\bibinfo {title} {Explicit {B}{C}{J} numerators from
  pure spinors},\ }\href {https://doi.org/10.1007/JHEP07(2011)092} {\bibfield
  {journal} {\bibinfo  {journal} {JHEP}\ }\textbf {\bibinfo {volume} {1107}},\
  \bibinfo {pages} {092}},\ \Eprint {https://arxiv.org/abs/1104.5224}
  {arXiv:1104.5224 [hep-th]} \BibitemShut {NoStop}%
\bibitem [{\citenamefont {Mafra}\ and\ \citenamefont
  {Schlotterer}(2014)}]{Mafra:2012kh}%
  \BibitemOpen
  \bibfield  {author} {\bibinfo {author} {\bibfnamefont {C.~R.}\ \bibnamefont
  {Mafra}}\ and\ \bibinfo {author} {\bibfnamefont {O.}~\bibnamefont
  {Schlotterer}},\ }\bibfield  {title} {\bibinfo {title} {The structure of
  $n$-point one-loop open superstring amplitudes},\ }\href
  {https://doi.org/10.1007/JHEP08(2014)099} {\bibfield  {journal} {\bibinfo
  {journal} {JHEP}\ }\textbf {\bibinfo {volume} {1408}},\ \bibinfo {pages}
  {099}},\ \Eprint {https://arxiv.org/abs/1203.6215} {arXiv:1203.6215 [hep-th]}
  \BibitemShut {NoStop}%
\bibitem [{\citenamefont {Fu}\ \emph {et~al.}(2013)\citenamefont {Fu},
  \citenamefont {Du},\ and\ \citenamefont {Feng}}]{Fu:2012uy}%
  \BibitemOpen
  \bibfield  {author} {\bibinfo {author} {\bibfnamefont {C.-H.}\ \bibnamefont
  {Fu}}, \bibinfo {author} {\bibfnamefont {Y.-J.}\ \bibnamefont {Du}},\ and\
  \bibinfo {author} {\bibfnamefont {B.}~\bibnamefont {Feng}},\ }\bibfield
  {title} {\bibinfo {title} {{An algebraic approach to {BCJ} numerators}},\
  }\href {https://doi.org/10.1007/JHEP03(2013)050} {\bibfield  {journal}
  {\bibinfo  {journal} {JHEP}\ }\textbf {\bibinfo {volume} {1303}},\ \bibinfo
  {pages} {050}},\ \Eprint {https://arxiv.org/abs/1212.6168} {arXiv:1212.6168
  [hep-th]} \BibitemShut {NoStop}%
\bibitem [{\citenamefont {Broedel}\ and\ \citenamefont
  {Dixon}(2012)}]{Broedel:2012rc}%
  \BibitemOpen
  \bibfield  {author} {\bibinfo {author} {\bibfnamefont {J.}~\bibnamefont
  {Broedel}}\ and\ \bibinfo {author} {\bibfnamefont {L.~J.}\ \bibnamefont
  {Dixon}},\ }\bibfield  {title} {\bibinfo {title} {Color--kinematics duality
  and double-copy construction for amplitudes from higher-dimension
  operators},\ }\href {https://doi.org/10.1007/JHEP10(1210)091} {\bibfield
  {journal} {\bibinfo  {journal} {JHEP}\ }\textbf {\bibinfo {volume} {2010}},\
  \bibinfo {pages} {091}},\ \Eprint {https://arxiv.org/abs/1208.0876}
  {arXiv:1208.0876 [hep-th]} \BibitemShut {NoStop}%
\bibitem [{\citenamefont {Broedel}\ \emph {et~al.}(2013)\citenamefont
  {Broedel}, \citenamefont {Schlotterer},\ and\ \citenamefont
  {Stieberger}}]{Broedel:2013tta}%
  \BibitemOpen
  \bibfield  {author} {\bibinfo {author} {\bibfnamefont {J.}~\bibnamefont
  {Broedel}}, \bibinfo {author} {\bibfnamefont {O.}~\bibnamefont
  {Schlotterer}},\ and\ \bibinfo {author} {\bibfnamefont {S.}~\bibnamefont
  {Stieberger}},\ }\bibfield  {title} {\bibinfo {title} {Polylogarithms,
  multiple zeta values and superstring amplitudes},\ }\href
  {https://doi.org/10.1002/prop.201300019} {\bibfield  {journal} {\bibinfo
  {journal} {Fortsch. Phys.}\ }\textbf {\bibinfo {volume} {61}},\ \bibinfo
  {pages} {812} (\bibinfo {year} {2013})},\ \Eprint
  {https://arxiv.org/abs/1304.7267} {arXiv:1304.7267 [hep-th]} \BibitemShut
  {NoStop}%
\bibitem [{\citenamefont {He}\ \emph {et~al.}(2016)\citenamefont {He},
  \citenamefont {Monteiro},\ and\ \citenamefont {Schlotterer}}]{He:2015wgf}%
  \BibitemOpen
  \bibfield  {author} {\bibinfo {author} {\bibfnamefont {S.}~\bibnamefont
  {He}}, \bibinfo {author} {\bibfnamefont {R.}~\bibnamefont {Monteiro}},\ and\
  \bibinfo {author} {\bibfnamefont {O.}~\bibnamefont {Schlotterer}},\
  }\bibfield  {title} {\bibinfo {title} {String-inspired {BCJ} numerators for
  one-loop {MHV} amplitudes},\ }\href {https://doi.org/10.1007/JHEP01(2016)171}
  {\bibfield  {journal} {\bibinfo  {journal} {JHEP}\ }\textbf {\bibinfo
  {volume} {1601}},\ \bibinfo {pages} {171}},\ \Eprint
  {https://arxiv.org/abs/1507.06288} {arXiv:1507.06288 [hep-th]} \BibitemShut
  {NoStop}%
\bibitem [{\citenamefont {Bjerrum-Bohr}\ \emph {et~al.}(2016)\citenamefont
  {Bjerrum-Bohr}, \citenamefont {Bourjaily}, \citenamefont {Damgaard},\ and\
  \citenamefont {Feng}}]{Bjerrum-Bohr:2016axv}%
  \BibitemOpen
  \bibfield  {author} {\bibinfo {author} {\bibfnamefont {N.~E.~J.}\
  \bibnamefont {Bjerrum-Bohr}}, \bibinfo {author} {\bibfnamefont {J.~L.}\
  \bibnamefont {Bourjaily}}, \bibinfo {author} {\bibfnamefont {P.~H.}\
  \bibnamefont {Damgaard}},\ and\ \bibinfo {author} {\bibfnamefont
  {B.}~\bibnamefont {Feng}},\ }\bibfield  {title} {\bibinfo {title}
  {Manifesting color--kinematics duality in the scattering equation
  formalism},\ }\href {https://doi.org/10.1007/JHEP09(2016)094} {\bibfield
  {journal} {\bibinfo  {journal} {JHEP}\ }\textbf {\bibinfo {volume} {1609}},\
  \bibinfo {pages} {094}},\ \Eprint {https://arxiv.org/abs/1608.00006}
  {arXiv:1608.00006 [hep-th]} \BibitemShut {NoStop}%
\bibitem [{\citenamefont {Mizera}(2020)}]{Mizera:2019blq}%
  \BibitemOpen
  \bibfield  {author} {\bibinfo {author} {\bibfnamefont {S.}~\bibnamefont
  {Mizera}},\ }\bibfield  {title} {\bibinfo {title} {Kinematic {J}acobi
  identity is a residue theorem: geometry of color--kinematics duality for
  gauge and gravity amplitudes},\ }\href
  {https://doi.org/10.1103/PhysRevLett.124.141601} {\bibfield  {journal}
  {\bibinfo  {journal} {Phys. Rev. Lett.}\ }\textbf {\bibinfo {volume} {124}},\
  \bibinfo {pages} {141601} (\bibinfo {year} {2020})},\ \Eprint
  {https://arxiv.org/abs/1912.03397} {arXiv:1912.03397 [hep-th]} \BibitemShut
  {NoStop}%
\bibitem [{\citenamefont {Chi}\ \emph {et~al.}(2022)\citenamefont {Chi},
  \citenamefont {Elvang}, \citenamefont {Herderschee}, \citenamefont {Jones},\
  and\ \citenamefont {Paranjape}}]{Chi:2021mio}%
  \BibitemOpen
  \bibfield  {author} {\bibinfo {author} {\bibfnamefont {H.-H.}\ \bibnamefont
  {Chi}}, \bibinfo {author} {\bibfnamefont {H.}~\bibnamefont {Elvang}},
  \bibinfo {author} {\bibfnamefont {A.}~\bibnamefont {Herderschee}}, \bibinfo
  {author} {\bibfnamefont {C.~R.~T.}\ \bibnamefont {Jones}},\ and\ \bibinfo
  {author} {\bibfnamefont {S.}~\bibnamefont {Paranjape}},\ }\bibfield  {title}
  {\bibinfo {title} {Generalizations of the double-copy: The {KLT} bootstrap},\
  }\href {https://doi.org/10.1007/JHEP03(2022)077} {\bibfield  {journal}
  {\bibinfo  {journal} {JHEP}\ }\textbf {\bibinfo {volume} {2203}},\ \bibinfo
  {pages} {077}},\ \Eprint {https://arxiv.org/abs/2106.12600} {arXiv:2106.12600
  [hep-th]} \BibitemShut {NoStop}%
\bibitem [{\citenamefont {Borsten}\ \emph {et~al.}()\citenamefont {Borsten},
  \citenamefont {Jur\v{c}o}, \citenamefont {Kim}, \citenamefont {Macrelli},
  \citenamefont {Saemann},\ and\ \citenamefont {Wolf}}]{Borsten:2023ned}%
  \BibitemOpen
  \bibfield  {author} {\bibinfo {author} {\bibfnamefont {L.}~\bibnamefont
  {Borsten}}, \bibinfo {author} {\bibfnamefont {B.}~\bibnamefont {Jur\v{c}o}},
  \bibinfo {author} {\bibfnamefont {H.}~\bibnamefont {Kim}}, \bibinfo {author}
  {\bibfnamefont {T.}~\bibnamefont {Macrelli}}, \bibinfo {author}
  {\bibfnamefont {C.}~\bibnamefont {Saemann}},\ and\ \bibinfo {author}
  {\bibfnamefont {M.}~\bibnamefont {Wolf}},\ }\bibfield  {title} {\bibinfo
  {title} {Double copy from tensor products of metric {BV}${}^{\color{gray}
  \blacksquare}$-algebras},\ }\href@noop {} {\ }\Eprint
  {https://arxiv.org/abs/2307.02563} {arXiv:2307.02563 [hep-th]} \BibitemShut
  {NoStop}%
\bibitem [{\citenamefont {Monteiro}\ and\ \citenamefont
  {O'Connell}(2011)}]{Monteiro:2011pc}%
  \BibitemOpen
  \bibfield  {author} {\bibinfo {author} {\bibfnamefont {R.}~\bibnamefont
  {Monteiro}}\ and\ \bibinfo {author} {\bibfnamefont {D.}~\bibnamefont
  {O'Connell}},\ }\bibfield  {title} {\bibinfo {title} {The kinematic algebra
  from the self-dual sector},\ }\href {https://doi.org/10.1007/JHEP07(2011)007}
  {\bibfield  {journal} {\bibinfo  {journal} {JHEP}\ }\textbf {\bibinfo
  {volume} {1107}},\ \bibinfo {pages} {007}},\ \Eprint
  {https://arxiv.org/abs/1105.2565} {arXiv:1105.2565 [hep-th]} \BibitemShut
  {NoStop}%
\bibitem [{\citenamefont {Bjerrum-Bohr}\ \emph {et~al.}(2012)\citenamefont
  {Bjerrum-Bohr}, \citenamefont {Damgaard}, \citenamefont {Monteiro},\ and\
  \citenamefont {O'Connell}}]{Bjerrum-Bohr:2012kaa}%
  \BibitemOpen
  \bibfield  {author} {\bibinfo {author} {\bibfnamefont {N.~E.~J.}\
  \bibnamefont {Bjerrum-Bohr}}, \bibinfo {author} {\bibfnamefont {P.~H.}\
  \bibnamefont {Damgaard}}, \bibinfo {author} {\bibfnamefont {R.}~\bibnamefont
  {Monteiro}},\ and\ \bibinfo {author} {\bibfnamefont {D.}~\bibnamefont
  {O'Connell}},\ }\bibfield  {title} {\bibinfo {title} {Algebras for
  amplitudes},\ }\href {https://doi.org/10.1007/JHEP06(2012)061} {\bibfield
  {journal} {\bibinfo  {journal} {JHEP}\ }\textbf {\bibinfo {volume} {1206}},\
  \bibinfo {pages} {061}},\ \Eprint {https://arxiv.org/abs/1203.0944}
  {arXiv:1203.0944 [hep-th]} \BibitemShut {NoStop}%
\bibitem [{\citenamefont {Boels}\ \emph {et~al.}(2013)\citenamefont {Boels},
  \citenamefont {Isermann}, \citenamefont {Monteiro},\ and\ \citenamefont
  {O'Connell}}]{Boels:2013bi}%
  \BibitemOpen
  \bibfield  {author} {\bibinfo {author} {\bibfnamefont {R.~H.}\ \bibnamefont
  {Boels}}, \bibinfo {author} {\bibfnamefont {R.~S.}\ \bibnamefont {Isermann}},
  \bibinfo {author} {\bibfnamefont {R.}~\bibnamefont {Monteiro}},\ and\
  \bibinfo {author} {\bibfnamefont {D.}~\bibnamefont {O'Connell}},\ }\bibfield
  {title} {\bibinfo {title} {Colour--kinematics duality for one-loop rational
  amplitudes},\ }\href {https://doi.org/10.1007/JHEP04(2013)107} {\bibfield
  {journal} {\bibinfo  {journal} {JHEP}\ }\textbf {\bibinfo {volume} {1304}},\
  \bibinfo {pages} {107}},\ \Eprint {https://arxiv.org/abs/1301.4165}
  {arXiv:1301.4165 [hep-th]} \BibitemShut {NoStop}%
\bibitem [{\citenamefont {Monteiro}\ and\ \citenamefont
  {O'Connell}(2014)}]{Monteiro:2013rya}%
  \BibitemOpen
  \bibfield  {author} {\bibinfo {author} {\bibfnamefont {R.}~\bibnamefont
  {Monteiro}}\ and\ \bibinfo {author} {\bibfnamefont {D.}~\bibnamefont
  {O'Connell}},\ }\bibfield  {title} {\bibinfo {title} {The kinematic algebras
  from the scattering equations},\ }\href
  {https://doi.org/10.1007/JHEP03(2014)110} {\bibfield  {journal} {\bibinfo
  {journal} {JHEP}\ }\textbf {\bibinfo {volume} {1403}},\ \bibinfo {pages}
  {110}},\ \Eprint {https://arxiv.org/abs/1311.1151} {arXiv:1311.1151 [hep-th]}
  \BibitemShut {NoStop}%
\bibitem [{\citenamefont {Fu}\ and\ \citenamefont
  {Krasnov}(2017)}]{Fu:2016plh}%
  \BibitemOpen
  \bibfield  {author} {\bibinfo {author} {\bibfnamefont {C.-H.}\ \bibnamefont
  {Fu}}\ and\ \bibinfo {author} {\bibfnamefont {K.}~\bibnamefont {Krasnov}},\
  }\bibfield  {title} {\bibinfo {title} {Colour--kinematics duality and the
  {D}rinfeld double of the {L}ie algebra of diffeomorphisms},\ }\href
  {https://doi.org/10.1007/JHEP01(2017)075} {\bibfield  {journal} {\bibinfo
  {journal} {JHEP}\ }\textbf {\bibinfo {volume} {1701}},\ \bibinfo {pages}
  {075}},\ \Eprint {https://arxiv.org/abs/1603.02033} {arXiv:1603.02033
  [hep-th]} \BibitemShut {NoStop}%
\bibitem [{\citenamefont {Monteiro}(2023)}]{Monteiro:2022lwm}%
  \BibitemOpen
  \bibfield  {author} {\bibinfo {author} {\bibfnamefont {R.}~\bibnamefont
  {Monteiro}},\ }\bibfield  {title} {\bibinfo {title} {Celestial chiral
  algebras, colour--kinematics duality and integrability},\ }\href
  {https://doi.org/10.1007/JHEP01(2023)092} {\bibfield  {journal} {\bibinfo
  {journal} {JHEP}\ }\textbf {\bibinfo {volume} {2301}},\ \bibinfo {pages}
  {092}},\ \Eprint {https://arxiv.org/abs/2208.11179} {arXiv:2208.11179
  [hep-th]} \BibitemShut {NoStop}%
\bibitem [{\citenamefont {Cheung}\ and\ \citenamefont
  {Shen}(2017)}]{Cheung:2016prv}%
  \BibitemOpen
  \bibfield  {author} {\bibinfo {author} {\bibfnamefont {C.}~\bibnamefont
  {Cheung}}\ and\ \bibinfo {author} {\bibfnamefont {C.-H.}\ \bibnamefont
  {Shen}},\ }\bibfield  {title} {\bibinfo {title} {Symmetry for
  flavor--kinematics duality from an action},\ }\href
  {https://doi.org/10.1103/PhysRevLett.118.121601} {\bibfield  {journal}
  {\bibinfo  {journal} {Phys. Rev. Lett.}\ }\textbf {\bibinfo {volume} {118}},\
  \bibinfo {pages} {121601} (\bibinfo {year} {2017})},\ \Eprint
  {https://arxiv.org/abs/1612.00868} {arXiv:1612.00868 [hep-th]} \BibitemShut
  {NoStop}%
\bibitem [{\citenamefont {Chen}\ \emph {et~al.}(2019)\citenamefont {Chen},
  \citenamefont {Johansson}, \citenamefont {Teng},\ and\ \citenamefont
  {Wang}}]{Chen:2019ywi}%
  \BibitemOpen
  \bibfield  {author} {\bibinfo {author} {\bibfnamefont {G.}~\bibnamefont
  {Chen}}, \bibinfo {author} {\bibfnamefont {H.}~\bibnamefont {Johansson}},
  \bibinfo {author} {\bibfnamefont {F.}~\bibnamefont {Teng}},\ and\ \bibinfo
  {author} {\bibfnamefont {T.}~\bibnamefont {Wang}},\ }\bibfield  {title}
  {\bibinfo {title} {On the kinematic algebra for {B}{C}{J} numerators beyond
  the {MHV} sector},\ }\href {https://doi.org/10.1007/JHEP11(2019)055}
  {\bibfield  {journal} {\bibinfo  {journal} {JHEP}\ }\textbf {\bibinfo
  {volume} {2019}},\ \bibinfo {pages} {055}},\ \Eprint
  {https://arxiv.org/abs/1906.10683} {arXiv:1906.10683 [hep-th]} \BibitemShut
  {NoStop}%
\bibitem [{\citenamefont {Chen}\ \emph {et~al.}(2021)\citenamefont {Chen},
  \citenamefont {Johansson}, \citenamefont {Teng},\ and\ \citenamefont
  {Wang}}]{Chen:2021chy}%
  \BibitemOpen
  \bibfield  {author} {\bibinfo {author} {\bibfnamefont {G.}~\bibnamefont
  {Chen}}, \bibinfo {author} {\bibfnamefont {H.}~\bibnamefont {Johansson}},
  \bibinfo {author} {\bibfnamefont {F.}~\bibnamefont {Teng}},\ and\ \bibinfo
  {author} {\bibfnamefont {T.}~\bibnamefont {Wang}},\ }\bibfield  {title}
  {\bibinfo {title} {Next-to-{MHV} {Y}ang--{M}ills kinematic algebra},\ }\href
  {https://doi.org/10.1007/JHEP10(2021)042} {\bibfield  {journal} {\bibinfo
  {journal} {JHEP}\ }\textbf {\bibinfo {volume} {2110}},\ \bibinfo {pages}
  {042}},\ \Eprint {https://arxiv.org/abs/2104.12726} {arXiv:2104.12726
  [hep-th]} \BibitemShut {NoStop}%
\bibitem [{\citenamefont {Brandhuber}\ \emph {et~al.}(2022)\citenamefont
  {Brandhuber}, \citenamefont {Chen}, \citenamefont {Johansson}, \citenamefont
  {Travaglini},\ and\ \citenamefont {Wen}}]{Brandhuber:2021bsf}%
  \BibitemOpen
  \bibfield  {author} {\bibinfo {author} {\bibfnamefont {A.}~\bibnamefont
  {Brandhuber}}, \bibinfo {author} {\bibfnamefont {G.}~\bibnamefont {Chen}},
  \bibinfo {author} {\bibfnamefont {H.}~\bibnamefont {Johansson}}, \bibinfo
  {author} {\bibfnamefont {G.}~\bibnamefont {Travaglini}},\ and\ \bibinfo
  {author} {\bibfnamefont {C.}~\bibnamefont {Wen}},\ }\bibfield  {title}
  {\bibinfo {title} {Kinematic {H}opf algebra for {BCJ} numerators in
  heavy-mass effective field theory and {Y}ang--{M}ills theory},\ }\href
  {https://doi.org/10.1103/PhysRevLett.128.121601} {\bibfield  {journal}
  {\bibinfo  {journal} {Phys. Rev. Lett.}\ }\textbf {\bibinfo {volume} {128}},\
  \bibinfo {pages} {121601} (\bibinfo {year} {2022})},\ \Eprint
  {https://arxiv.org/abs/2111.15649} {arXiv:2111.15649 [hep-th]} \BibitemShut
  {NoStop}%
\bibitem [{\citenamefont {Cheung}\ and\ \citenamefont
  {Mangan}(2021)}]{Cheung:2021zvb}%
  \BibitemOpen
  \bibfield  {author} {\bibinfo {author} {\bibfnamefont {C.}~\bibnamefont
  {Cheung}}\ and\ \bibinfo {author} {\bibfnamefont {J.}~\bibnamefont
  {Mangan}},\ }\bibfield  {title} {\bibinfo {title} {Covariant
  color--kinematics duality},\ }\href {https://doi.org/10.1007/JHEP11(2021)069}
  {\bibfield  {journal} {\bibinfo  {journal} {JHEP}\ }\textbf {\bibinfo
  {volume} {2111}},\ \bibinfo {pages} {069}},\ \Eprint
  {https://arxiv.org/abs/2108.02276} {arXiv:2108.02276 [hep-th]} \BibitemShut
  {NoStop}%
\bibitem [{\citenamefont {Ben-Shahar}\ and\ \citenamefont
  {Guillen}(2021)}]{Ben-Shahar:2021doh}%
  \BibitemOpen
  \bibfield  {author} {\bibinfo {author} {\bibfnamefont {M.}~\bibnamefont
  {Ben-Shahar}}\ and\ \bibinfo {author} {\bibfnamefont {M.}~\bibnamefont
  {Guillen}},\ }\bibfield  {title} {\bibinfo {title} {10d super-{Y}ang--{M}ills
  scattering amplitudes from its pure spinor action},\ }\href
  {https://doi.org/10.1007/JHEP12(2021)014} {\bibfield  {journal} {\bibinfo
  {journal} {JHEP}\ }\textbf {\bibinfo {volume} {2112}},\ \bibinfo {pages}
  {014}},\ \Eprint {https://arxiv.org/abs/2108.11708} {arXiv:2108.11708
  [hep-th]} \BibitemShut {NoStop}%
\bibitem [{\citenamefont {Bern}\ \emph
  {et~al.}(2010{\natexlab{b}})\citenamefont {Bern}, \citenamefont {Dennen},
  \citenamefont {Huang},\ and\ \citenamefont {Kiermaier}}]{Bern:2010yg}%
  \BibitemOpen
  \bibfield  {author} {\bibinfo {author} {\bibfnamefont {Z.}~\bibnamefont
  {Bern}}, \bibinfo {author} {\bibfnamefont {T.}~\bibnamefont {Dennen}},
  \bibinfo {author} {\bibfnamefont {Y.-t.}\ \bibnamefont {Huang}},\ and\
  \bibinfo {author} {\bibfnamefont {M.}~\bibnamefont {Kiermaier}},\ }\bibfield
  {title} {\bibinfo {title} {Gravity as the square of gauge theory},\ }\href
  {https://doi.org/10.1103/PhysRevD.82.065003} {\bibfield  {journal} {\bibinfo
  {journal} {Phys. Rev. D}\ }\textbf {\bibinfo {volume} {82}},\ \bibinfo
  {pages} {065003} (\bibinfo {year} {2010}{\natexlab{b}})},\ \Eprint
  {https://arxiv.org/abs/1004.0693} {arXiv:1004.0693 [hep-th]} \BibitemShut
  {NoStop}%
\bibitem [{\citenamefont {Borsten}\ \emph
  {et~al.}(2021{\natexlab{a}})\citenamefont {Borsten}, \citenamefont
  {Jur\v{c}o}, \citenamefont {Kim}, \citenamefont {Macrelli}, \citenamefont
  {Saemann},\ and\ \citenamefont {Wolf}}]{Borsten:2020zgj}%
  \BibitemOpen
  \bibfield  {author} {\bibinfo {author} {\bibfnamefont {L.}~\bibnamefont
  {Borsten}}, \bibinfo {author} {\bibfnamefont {B.}~\bibnamefont {Jur\v{c}o}},
  \bibinfo {author} {\bibfnamefont {H.}~\bibnamefont {Kim}}, \bibinfo {author}
  {\bibfnamefont {T.}~\bibnamefont {Macrelli}}, \bibinfo {author}
  {\bibfnamefont {C.}~\bibnamefont {Saemann}},\ and\ \bibinfo {author}
  {\bibfnamefont {M.}~\bibnamefont {Wolf}},\ }\bibfield  {title} {\bibinfo
  {title} {{BRST}--{L}agrangian double copy of {Y}ang--{M}ills theory},\ }\href
  {https://doi.org/10.1103/PhysRevLett.126.191601} {\bibfield  {journal}
  {\bibinfo  {journal} {Phys. Rev. Lett.}\ }\textbf {\bibinfo {volume} {126}},\
  \bibinfo {pages} {191601} (\bibinfo {year} {2021}{\natexlab{a}})},\ \Eprint
  {https://arxiv.org/abs/2007.13803} {arXiv:2007.13803 [hep-th]} \BibitemShut
  {NoStop}%
\bibitem [{\citenamefont {Borsten}\ \emph
  {et~al.}(2021{\natexlab{b}})\citenamefont {Borsten}, \citenamefont
  {Jur\v{c}o}, \citenamefont {Kim}, \citenamefont {Macrelli}, \citenamefont
  {Saemann},\ and\ \citenamefont {Wolf}}]{Borsten:2021hua}%
  \BibitemOpen
  \bibfield  {author} {\bibinfo {author} {\bibfnamefont {L.}~\bibnamefont
  {Borsten}}, \bibinfo {author} {\bibfnamefont {B.}~\bibnamefont {Jur\v{c}o}},
  \bibinfo {author} {\bibfnamefont {H.}~\bibnamefont {Kim}}, \bibinfo {author}
  {\bibfnamefont {T.}~\bibnamefont {Macrelli}}, \bibinfo {author}
  {\bibfnamefont {C.}~\bibnamefont {Saemann}},\ and\ \bibinfo {author}
  {\bibfnamefont {M.}~\bibnamefont {Wolf}},\ }\bibfield  {title} {\bibinfo
  {title} {{Double copy from homotopy algebras}},\ }\href
  {https://doi.org/10.1002/prop.202100075} {\bibfield  {journal} {\bibinfo
  {journal} {Fortsch. Phys.}\ }\textbf {\bibinfo {volume} {69}},\ \bibinfo
  {pages} {2100075} (\bibinfo {year} {2021}{\natexlab{b}})},\ \Eprint
  {https://arxiv.org/abs/2102.11390} {arXiv:2102.11390 [hep-th]} \BibitemShut
  {NoStop}%
\bibitem [{\citenamefont {Ben-Shahar}\ and\ \citenamefont
  {Johansson}(2022)}]{Ben-Shahar:2021zww}%
  \BibitemOpen
  \bibfield  {author} {\bibinfo {author} {\bibfnamefont {M.}~\bibnamefont
  {Ben-Shahar}}\ and\ \bibinfo {author} {\bibfnamefont {H.}~\bibnamefont
  {Johansson}},\ }\bibfield  {title} {\bibinfo {title} {Off-shell
  color--kinematics duality for {C}hern--{S}imons},\ }\href
  {https://doi.org/10.1007/JHEP08(2022)035} {\bibfield  {journal} {\bibinfo
  {journal} {JHEP}\ }\textbf {\bibinfo {volume} {2208}},\ \bibinfo {pages}
  {035}},\ \Eprint {https://arxiv.org/abs/2112.11452} {arXiv:2112.11452
  [hep-th]} \BibitemShut {NoStop}%
\bibitem [{\citenamefont {Costello}(2021)}]{Costello:2021bah}%
  \BibitemOpen
  \bibfield  {author} {\bibinfo {author} {\bibfnamefont {K.~J.}\ \bibnamefont
  {Costello}},\ }\bibfield  {title} {\bibinfo {title} {Quantizing local
  holomorphic field theories on twistor space},\ }\href@noop {} {\  (\bibinfo
  {year} {2021})},\ \Eprint {https://arxiv.org/abs/2111.08879}
  {arXiv:2111.08879 [hep-th]} \BibitemShut {NoStop}%
\bibitem [{\citenamefont {Batalin}\ and\ \citenamefont
  {Vilkovisky}(1981)}]{Batalin:1981jr}%
  \BibitemOpen
  \bibfield  {author} {\bibinfo {author} {\bibfnamefont {I.~A.}\ \bibnamefont
  {Batalin}}\ and\ \bibinfo {author} {\bibfnamefont {G.~A.}\ \bibnamefont
  {Vilkovisky}},\ }\bibfield  {title} {\bibinfo {title} {{Gauge algebra and
  quantization}},\ }\bibfield  {booktitle} {\emph {\bibinfo {booktitle}
  {{Second Seminar on Quantum Gravity Moscow, USSR, October 13-15, 1981}}},\
  }\href {https://doi.org/10.1016/0370-2693(81)90205-7} {\bibfield  {journal}
  {\bibinfo  {journal} {Phys. Lett. B}\ }\textbf {\bibinfo {volume} {102}},\
  \bibinfo {pages} {27} (\bibinfo {year} {1981})}\BibitemShut {NoStop}%
\bibitem [{\citenamefont {Schwarz}(1993)}]{Schwarz:1992nx}%
  \BibitemOpen
  \bibfield  {author} {\bibinfo {author} {\bibfnamefont {A.}~\bibnamefont
  {Schwarz}},\ }\bibfield  {title} {\bibinfo {title} {{Geometry of
  Batalin--Vilkovisky quantization}},\ }\href
  {https://doi.org/10.1007/BF02097392} {\bibfield  {journal} {\bibinfo
  {journal} {Commun. Math. Phys.}\ }\textbf {\bibinfo {volume} {155}},\
  \bibinfo {pages} {249} (\bibinfo {year} {1993})},\ \Eprint
  {https://arxiv.org/abs/hep-th/9205088} {arXiv:hep-th/9205088} \BibitemShut
  {NoStop}%
\bibitem [{\citenamefont {Jur\v{c}o}\ \emph
  {et~al.}(2019{\natexlab{a}})\citenamefont {Jur\v{c}o}, \citenamefont
  {Raspollini}, \citenamefont {Saemann},\ and\ \citenamefont
  {Wolf}}]{Jurco:2018sby}%
  \BibitemOpen
  \bibfield  {author} {\bibinfo {author} {\bibfnamefont {B.}~\bibnamefont
  {Jur\v{c}o}}, \bibinfo {author} {\bibfnamefont {L.}~\bibnamefont
  {Raspollini}}, \bibinfo {author} {\bibfnamefont {C.}~\bibnamefont
  {Saemann}},\ and\ \bibinfo {author} {\bibfnamefont {M.}~\bibnamefont
  {Wolf}},\ }\bibfield  {title} {\bibinfo {title} {{$L_\infty$-algebras of
  classical field theories and the {B}atalin--{V}ilkovisky formalism}},\ }\href
  {https://doi.org/10.1002/prop.201900025} {\bibfield  {journal} {\bibinfo
  {journal} {Fortsch. Phys.}\ }\textbf {\bibinfo {volume} {67}},\ \bibinfo
  {pages} {1900025} (\bibinfo {year} {2019}{\natexlab{a}})},\ \Eprint
  {https://arxiv.org/abs/1809.09899} {arXiv:1809.09899 [hep-th]} \BibitemShut
  {NoStop}%
\bibitem [{\citenamefont {Koszul}(1985)}]{koszul1985crochet}%
  \BibitemOpen
  \bibfield  {author} {\bibinfo {author} {\bibfnamefont {J.-L.}\ \bibnamefont
  {Koszul}},\ }\bibfield  {title} {\bibinfo {title} {Crochet de
  {S}chouten-{N}ijenhuis et cohomologie},\ }\href
  {http://www.numdam.org/item?id=AST_1985__S131__257_0} {\bibfield  {journal}
  {\bibinfo  {journal} {Ast{\'e}risque}\ }\textbf {\bibinfo {volume} {131}},\
  \bibinfo {pages} {257} (\bibinfo {year} {1985})}\BibitemShut {NoStop}%
\bibitem [{\citenamefont {Coll}\ and\ \citenamefont
  {Ferrando}(2004)}]{Coll:2003ym}%
  \BibitemOpen
  \bibfield  {author} {\bibinfo {author} {\bibfnamefont {B.}~\bibnamefont
  {Coll}}\ and\ \bibinfo {author} {\bibfnamefont {J.~J.}\ \bibnamefont
  {Ferrando}},\ }\bibfield  {title} {\bibinfo {title} {On the {L}eibniz
  bracket, the {S}chouten bracket and the {L}aplacian},\ }\href
  {https://doi.org/10.1063/1.1738188} {\bibfield  {journal} {\bibinfo
  {journal} {J. Math. Phys.}\ }\textbf {\bibinfo {volume} {45}},\ \bibinfo
  {pages} {2405} (\bibinfo {year} {2004})},\ \Eprint
  {https://arxiv.org/abs/gr-qc/0306102} {arXiv:gr-qc/0306102 [gr-qc]}
  \BibitemShut {NoStop}%
\bibitem [{\citenamefont {Jur\v{c}o}\ \emph
  {et~al.}(2019{\natexlab{b}})\citenamefont {Jur\v{c}o}, \citenamefont
  {Macrelli}, \citenamefont {Raspollini}, \citenamefont {Saemann},\ and\
  \citenamefont {Wolf}}]{Jurco:2019bvp}%
  \BibitemOpen
  \bibfield  {author} {\bibinfo {author} {\bibfnamefont {B.}~\bibnamefont
  {Jur\v{c}o}}, \bibinfo {author} {\bibfnamefont {T.}~\bibnamefont {Macrelli}},
  \bibinfo {author} {\bibfnamefont {L.}~\bibnamefont {Raspollini}}, \bibinfo
  {author} {\bibfnamefont {C.}~\bibnamefont {Saemann}},\ and\ \bibinfo {author}
  {\bibfnamefont {M.}~\bibnamefont {Wolf}},\ }\bibfield  {title} {\bibinfo
  {title} {{$L_\infty$-algebras, the BV formalism, and classical fields}},\
  }\href@noop {} {\  (\bibinfo {year} {2019}{\natexlab{b}})},\ \bibinfo {note}
  {in: ``Higher Structures in M-Theory,'' proceedings of the
  \href{http://www.maths.dur.ac.uk/lms/109/index.html}{LMS/EPSRC Durham
  Symposium}, 12--18 August 2018},\ \Eprint {https://arxiv.org/abs/1903.02887}
  {arXiv:1903.02887 [hep-th]} \BibitemShut {NoStop}%
\bibitem [{\citenamefont {Markl}(2001)}]{Markl:0002130}%
  \BibitemOpen
  \bibfield  {author} {\bibinfo {author} {\bibfnamefont {M.}~\bibnamefont
  {Markl}},\ }\bibfield  {title} {\bibinfo {title} {Ideal perturbation lemma},\
  }\href {https://doi.org/10.1081/agb-100106814} {\bibfield  {journal}
  {\bibinfo  {journal} {Commun. Algebra}\ }\textbf {\bibinfo {volume} {29}},\
  \bibinfo {pages} {5209} (\bibinfo {year} {2001})},\ \Eprint
  {https://arxiv.org/abs/math.AT/0002130} {arXiv:math.AT/0002130 [math.AT]}
  \BibitemShut {NoStop}%
\bibitem [{\citenamefont {Getzler}(1994)}]{Getzler:1994yd}%
  \BibitemOpen
  \bibfield  {author} {\bibinfo {author} {\bibfnamefont {E.}~\bibnamefont
  {Getzler}},\ }\bibfield  {title} {\bibinfo {title} {{Batalin--Vilkovisky
  algebras and two-dimensional topological field theories}},\ }\href
  {https://doi.org/10.1007/BF02102639} {\bibfield  {journal} {\bibinfo
  {journal} {Commun. Math. Phys.}\ }\textbf {\bibinfo {volume} {159}},\
  \bibinfo {pages} {265} (\bibinfo {year} {1994})},\ \Eprint
  {https://arxiv.org/abs/hep-th/9212043} {arXiv:hep-th/9212043 [hep-th]}
  \BibitemShut {NoStop}%
\bibitem [{\citenamefont {Popov}\ and\ \citenamefont
  {Saemann}(2005)}]{Popov:2004rb}%
  \BibitemOpen
  \bibfield  {author} {\bibinfo {author} {\bibfnamefont {A.~D.}\ \bibnamefont
  {Popov}}\ and\ \bibinfo {author} {\bibfnamefont {C.}~\bibnamefont
  {Saemann}},\ }\bibfield  {title} {\bibinfo {title} {On supertwistors, the
  {P}enrose--{W}ard transform and {$\caN = 4$} super {Y}ang--{M}ills theory},\
  }\href {https://doi.org/10.4310/ATMP.2005.v9.n6.a2} {\bibfield  {journal}
  {\bibinfo  {journal} {Adv. Theor. Math. Phys.}\ }\textbf {\bibinfo {volume}
  {9}},\ \bibinfo {pages} {931} (\bibinfo {year} {2005})},\ \Eprint
  {https://arxiv.org/abs/hep-th/0405123} {arXiv:hep-th/0405123} \BibitemShut
  {NoStop}%
\bibitem [{\citenamefont {Wolf}(2010)}]{Wolf:2010av}%
  \BibitemOpen
  \bibfield  {author} {\bibinfo {author} {\bibfnamefont {M.}~\bibnamefont
  {Wolf}},\ }\bibfield  {title} {\bibinfo {title} {{A first course on twistors,
  integrability and gluon scattering amplitudes}},\ }\href
  {https://doi.org/10.1088/1751-8113/43/39/393001} {\bibfield  {journal}
  {\bibinfo  {journal} {J. Phys. A}\ }\textbf {\bibinfo {volume} {43}},\
  \bibinfo {pages} {393001} (\bibinfo {year} {2010})},\ \Eprint
  {https://arxiv.org/abs/1001.3871} {arXiv:1001.3871 [hep-th]} \BibitemShut
  {NoStop}%
\bibitem [{\citenamefont {Mason}(2005)}]{Mason:2005zm}%
  \BibitemOpen
  \bibfield  {author} {\bibinfo {author} {\bibfnamefont {L.~J.}\ \bibnamefont
  {Mason}},\ }\bibfield  {title} {\bibinfo {title} {Twistor actions for
  non-self-dual fields; a new foundation for twistor-string theory},\ }\href
  {https://doi.org/10.1088/1126-6708/2005/10/009} {\bibfield  {journal}
  {\bibinfo  {journal} {JHEP}\ }\textbf {\bibinfo {volume} {0510}},\ \bibinfo
  {pages} {009}},\ \Eprint {https://arxiv.org/abs/hep-th/0507269}
  {arXiv:hep-th/0507269} \BibitemShut {NoStop}%
\bibitem [{\citenamefont {Boels}\ \emph {et~al.}(2007)\citenamefont {Boels},
  \citenamefont {Mason},\ and\ \citenamefont {Skinner}}]{Boels:2006ir}%
  \BibitemOpen
  \bibfield  {author} {\bibinfo {author} {\bibfnamefont {R.}~\bibnamefont
  {Boels}}, \bibinfo {author} {\bibfnamefont {L.}~\bibnamefont {Mason}},\ and\
  \bibinfo {author} {\bibfnamefont {D.}~\bibnamefont {Skinner}},\ }\bibfield
  {title} {\bibinfo {title} {Supersymmetric gauge theories in twistor space},\
  }\href {https://doi.org/10.1088/1126-6708/2007/02/014} {\bibfield  {journal}
  {\bibinfo  {journal} {JHEP}\ }\textbf {\bibinfo {volume} {0702}},\ \bibinfo
  {pages} {014}},\ \Eprint {https://arxiv.org/abs/hep-th/0604040}
  {arXiv:hep-th/0604040} \BibitemShut {NoStop}%
\bibitem [{\citenamefont {Siegel}(1992)}]{Siegel:1992xp}%
  \BibitemOpen
  \bibfield  {author} {\bibinfo {author} {\bibfnamefont {W.}~\bibnamefont
  {Siegel}},\ }\bibfield  {title} {\bibinfo {title} {{$N=2$} (4) string theory
  is self-dual {$N=4$} {Y}ang--{M}ills theory},\ }\href
  {https://doi.org/10.1103/PhysRevD.46.R3235} {\bibfield  {journal} {\bibinfo
  {journal} {Phys. Rev.}\ }\textbf {\bibinfo {volume} {D46}},\ \bibinfo {pages}
  {3235} (\bibinfo {year} {1992})},\ \Eprint
  {https://arxiv.org/abs/hep-th/9205075} {arXiv:hep-th/9205075 [hep-th]}
  \BibitemShut {NoStop}%
\bibitem [{\citenamefont {Witten}(2004)}]{Witten:2003nn}%
  \BibitemOpen
  \bibfield  {author} {\bibinfo {author} {\bibfnamefont {E.}~\bibnamefont
  {Witten}},\ }\bibfield  {title} {\bibinfo {title} {{Perturbative gauge theory
  as a string theory in twistor space}},\ }\href
  {https://doi.org/10.1007/s00220-004-1187-3} {\bibfield  {journal} {\bibinfo
  {journal} {Commun. Math. Phys.}\ }\textbf {\bibinfo {volume} {252}},\
  \bibinfo {pages} {189} (\bibinfo {year} {2004})},\ \Eprint
  {https://arxiv.org/abs/hep-th/0312171} {arXiv:hep-th/0312171 [hep-th]}
  \BibitemShut {NoStop}%
\bibitem [{\citenamefont {Saemann}(2005)}]{Saemann:2004tt}%
  \BibitemOpen
  \bibfield  {author} {\bibinfo {author} {\bibfnamefont {C.}~\bibnamefont
  {Saemann}},\ }\bibfield  {title} {\bibinfo {title} {The topological {B}-model
  on fattened complex manifolds and subsectors of {$\caN = 4$} self-dual
  {Y}ang--{M}ills theory},\ }\href
  {https://doi.org/10.1088/1126-6708/2005/01/042} {\bibfield  {journal}
  {\bibinfo  {journal} {JHEP}\ }\textbf {\bibinfo {volume} {0501}},\ \bibinfo
  {pages} {042}},\ \Eprint {https://arxiv.org/abs/hep-th/0410292}
  {hep-th/0410292} \BibitemShut {NoStop}%
\bibitem [{\citenamefont {Popov}\ and\ \citenamefont
  {Wolf}(2004)}]{Popov:2004nk}%
  \BibitemOpen
  \bibfield  {author} {\bibinfo {author} {\bibfnamefont {A.~D.}\ \bibnamefont
  {Popov}}\ and\ \bibinfo {author} {\bibfnamefont {M.}~\bibnamefont {Wolf}},\
  }\bibfield  {title} {\bibinfo {title} {Topological b-model on weighted
  projective spaces and self-dual models in four dimensions},\ }\href
  {https://doi.org/10.1088/1126-6708/2004/09/007} {\bibfield  {journal}
  {\bibinfo  {journal} {JHEP}\ }\textbf {\bibinfo {volume} {0409}},\ \bibinfo
  {pages} {007}},\ \Eprint {https://arxiv.org/abs/hep-th/0406224}
  {arXiv:hep-th/0406224} \BibitemShut {NoStop}%
\bibitem [{\citenamefont {Popov}\ \emph {et~al.}(2005)\citenamefont {Popov},
  \citenamefont {Saemann},\ and\ \citenamefont {Wolf}}]{Popov:2005uv}%
  \BibitemOpen
  \bibfield  {author} {\bibinfo {author} {\bibfnamefont {A.~D.}\ \bibnamefont
  {Popov}}, \bibinfo {author} {\bibfnamefont {C.}~\bibnamefont {Saemann}},\
  and\ \bibinfo {author} {\bibfnamefont {M.}~\bibnamefont {Wolf}},\ }\bibfield
  {title} {\bibinfo {title} {{The topological {B}-model on a mini-supertwistor
  space and supersymmetric {B}ogomolny monopole equations}},\ }\href
  {https://doi.org/10.1088/1126-6708/2005/10/058} {\bibfield  {journal}
  {\bibinfo  {journal} {JHEP}\ }\textbf {\bibinfo {volume} {0510}},\ \bibinfo
  {pages} {058}},\ \Eprint {https://arxiv.org/abs/hep-th/0505161}
  {arXiv:hep-th/0505161} \BibitemShut {NoStop}%
\bibitem [{\citenamefont {Boggess}(1991)}]{Boggess:1991aa}%
  \BibitemOpen
  \bibfield  {author} {\bibinfo {author} {\bibfnamefont {A.}~\bibnamefont
  {Boggess}},\ }\href {https://doi.org/10.1201/9781315140445} {\emph {\bibinfo
  {title} {CR manifolds and the tangential {C}auchy--{R}iemann complex}}}\
  (\bibinfo  {publisher} {CRC press},\ \bibinfo {year} {1991})\BibitemShut
  {NoStop}%
\bibitem [{\citenamefont {Movshev}(2004)}]{Movshev:2004ub}%
  \BibitemOpen
  \bibfield  {author} {\bibinfo {author} {\bibfnamefont {M.}~\bibnamefont
  {Movshev}},\ }\bibfield  {title} {\bibinfo {title} {{Y}ang--{M}ills theory
  and a superquadric},\ }\href@noop {} {\  (\bibinfo {year} {2004})},\ \bibinfo
  {note} {in: ``Algebra, arithmetic, and geometry. Volume II. In honor of Yu.
  I. Manin,'' eds. Yuri Tschinkel and Yuri Zarhin, Progress in Mathematics
  (vol. 270), Birkhäuser 2009},\ \Eprint
  {https://arxiv.org/abs/hep-th/0411111} {arXiv:hep-th/0411111 [hep-th]}
  \BibitemShut {NoStop}%
\bibitem [{\citenamefont {Mason}\ and\ \citenamefont
  {Skinner}(2006)}]{Mason:2005kn}%
  \BibitemOpen
  \bibfield  {author} {\bibinfo {author} {\bibfnamefont {L.~J.}\ \bibnamefont
  {Mason}}\ and\ \bibinfo {author} {\bibfnamefont {D.}~\bibnamefont
  {Skinner}},\ }\bibfield  {title} {\bibinfo {title} {{An ambitwistor
  {Y}ang--{M}ills Lagrangian}},\ }\href
  {https://doi.org/10.1016/j.physletb.2006.02.061} {\bibfield  {journal}
  {\bibinfo  {journal} {Phys. Lett. B}\ }\textbf {\bibinfo {volume} {636}},\
  \bibinfo {pages} {60} (\bibinfo {year} {2006})},\ \Eprint
  {https://arxiv.org/abs/hep-th/0510262} {hep-th/0510262} \BibitemShut
  {NoStop}%
\bibitem [{\citenamefont {Bern}\ \emph {et~al.}(2016)\citenamefont {Bern},
  \citenamefont {Davies},\ and\ \citenamefont {Nohle}}]{Bern:2015ooa}%
  \BibitemOpen
  \bibfield  {author} {\bibinfo {author} {\bibfnamefont {Z.}~\bibnamefont
  {Bern}}, \bibinfo {author} {\bibfnamefont {S.}~\bibnamefont {Davies}},\ and\
  \bibinfo {author} {\bibfnamefont {J.}~\bibnamefont {Nohle}},\ }\bibfield
  {title} {\bibinfo {title} {Double-copy constructions and unitarity cuts},\
  }\href {https://doi.org/10.1103/PhysRevD.93.105015} {\bibfield  {journal}
  {\bibinfo  {journal} {Phys. Rev. D}\ }\textbf {\bibinfo {volume} {93}},\
  \bibinfo {pages} {105015} (\bibinfo {year} {2016})},\ \Eprint
  {https://arxiv.org/abs/1510.03448} {arXiv:1510.03448 [hep-th]} \BibitemShut
  {NoStop}%
\end{thebibliography}%

\end{document}